\documentclass[12pt]{article}

\usepackage{jheppub,amsbsy,amssymb,latexsym,amsfonts,amsmath,epsfig,multicol,bbm}

\allowdisplaybreaks

\title{On the classical equivalence of monodromy matrices in squashed sigma model}

\author{Io Kawaguchi$^{\ast}$\footnote{E-mail:~io@gauge.scphys.kyoto-u.ac.jp}, 
Takuya Matsumoto$^{\dagger,\ddagger}$\footnote{E-mail:~tmatsumoto@usyd.edu.au} 
and Kentaroh Yoshida$^{\ast}$\footnote{E-mail:~kyoshida@gauge.scphys.kyoto-u.ac.jp}}
\affiliation{$^{\ast}${\it Department of Physics, Kyoto University Kyoto 606-8502, Japan} \\
$^{\dagger}${\it School of Mathematics and Statistics, University of Sydney, NSW 2006, Australia} \\
$^{\ddagger}${\it Graduate School of Mathematics, Nagoya University, Nagoya 464-8602, Japan} }

\arxivnumber{1203.3400[hep-th]}

\abstract{
We proceed to study the hybrid integrable structure in two-dimensional non-linear sigma models 
with target space three-dimensional squashed spheres. A quantum affine algebra and a pair of Yangian 
algebras are realized in the sigma models and, according to them, there are two descriptions 
to describe the classical dynamics 1) the trigonometric description and 2) the rational description, 
respectively. For every description, a Lax pair is constructed and the associated monodromy matrix 
is also constructed. In this paper we show the gauge-equivalence of the monodromy matrices 
in the trigonometric and rational description under a certain relation between spectral parameters 
and the rescalings of $sl(2)$ generators. 
}


\begin{document}

\maketitle

\section{Introduction}
It is no exaggeration to say that the most fascinating topic in string theory is the AdS/CFT correspondence 
\cite{Maldacena}. It provides a specific approach to quantum gravity as well as a useful tool to study 
strongly-coupled systems. An enormous amount of evidence support the duality, but still there is no rigorous 
proof of it. One may attempt to ask what mechanism is responsible for AdS/CFT. At the present time, 
the integrability is recognized as the fundamental structure of AdS/CFT  
(For a comprehensive review, see \cite{review}).

\medskip 

The next issue is to consider integrable deformations of AdS/CFT. In this direction there are preceding works 
such as $\beta$-deformation \cite{LS} and its gravity dual \cite{LM,Frolov,Berenstein}\,, 
and $q$-deformation of the world-sheet S-matrix \cite{BK,BGM,Hollowood}. 
Apart from them, we are interested in three-dimensional squashed spheres and warped AdS$_3$\,. 
These geometries appear in recent studies like holographic condensed matter \cite{Kraus}, Kerr/CFT \cite{Kerr/CFT} 
and warped AdS$_3$/dipole CFT$_2$ \cite{Guica,SS}. The potential applications to these topics 
make it significant to study the integrable structure of two-dimensional non-linear sigma models 
with target space warped AdS$_3$ and squashed spheres.  

\medskip 

In this paper we concentrate on the classical integrable structure of sigma models with squashed spheres. 
The reason is that warped AdS$_3$ geometries are obtained via double Wick rotations of squashed spheres 
and the classical analysis performed here is valid irrespective of compactness of target space. 
We refer to the sigma models as ``the squashed sigma models'' as an abbreviation hereafter.  

\medskip 

In a series of works \cite{KY,KOY,KYhybrid,KMY} (For a short summary see \cite{KY-summary}), 
we have shown that quantum affine algebra and Yangian algebra are realized in the squashed sigma models\footnote{
The classical integrability is discussed also from T-duality argument \cite{ORU}.}. 
According to them, there are two descriptions to describe the classical dynamics: 
1) the rational description and 2) the trigonometric description. Depending on the description, 
two kinds of Lax pair, which lead to the identical classical equations of motion, are constructed and 
also there are the corresponding monodromy matrices. 

\medskip 

This means the ``local'' equivalence 
of the two descriptions and does not imply the equivalence of classical moduli spaces, namely ``global'' equivalence. 
In other words, the ``global'' equivalence is equivalent to the left-right symmetry. 
In fact, the ``local'' equivalence has been well known, while it has been believed that the ``global'' equivalence 
is not realized because the universality class of Lax pairs (i.e., topology of classical moduli space), 
spectral parameters and the number of poles are different between the two descriptions\footnote{
For example, see the sentence just below (2.18) of \cite{Forgacs:2000eu}\,. 
We are grateful to Adam Rej for drawing our attention to this article.}. 

\medskip 

We proceed here to study the classical integrable structure of squashed sigma models. 
We show the gauge-equivalence of monodromy matrices in the trigonometric and rational descriptions 
under the relation of spectral parameters and the rescalings of $sl(2)$ generators. 
As a result, the trigonometric description is shown to be equivalent to a composite of 
the rational descriptions. 
That is, the ``global'' equivalence is accurately realized 
even after squashing the target space geometry, in contrast to the folklore 
which has been believed so far without concrete proof.  
All of the difficulties mentioned in the previous paragraph are resolved 
by taking account of the two rational descriptions 
and finding out the relation between spectral parameters. 
Moreover, we find the ``reduced'' trigonometric description 
that works as the Lax pair at least at the classical level. With this description, the equivalence of 
the monodromy matrices becomes very apparent. 

\medskip 

This paper is organized as follows. In section 2 we introduce the squashed sigma models and 
the monodromy matrices in the trigonometric and rational descriptions. In section 3 
the monodromy matrices are expanded around some points and the relation of spectral parameters is 
deduced. 
In section 4 we show the gauge-equivalence of monodromy matrices under the spectral parameter relation 
and the rescalings of $sl(2)$ generators. The reducibility of the trigonometric Lax pair is also discussed.  
Section 5 is devoted to conclusion and discussion.

\section{Preliminaries}

We introduce the classical action of squashed sigma models 
and give a short review on a series of works \cite{KY,KOY,KYhybrid,KMY,KY-summary}, 
including some new results. Two descriptions to describe the classical dynamics 
are explained with monodromy matrices, which will be the main objects in the following discussion.

\subsection{The classical action of squashed sigma models}

First of all, let us introduce the $su(2)$ Lie algebra generators $T^{a}~(a=1,2,3)$ satisfying  
\begin{eqnarray}
\left[T^{a},T^{b}\right] = \varepsilon^{ab}_{~~c}T^{c}\,, \quad 
{\rm Tr}\left(T^{a}T^{b}\right) = -\frac{1}{2}\delta^{ab}\,. \nonumber 
\end{eqnarray}
The totally antisymmetric tensor $\varepsilon^{ab}_{~~c}$ is normalized as $\varepsilon_{123}=+1$\,. 

\medskip 

By using the left-invariant one-form, 
\[
J \equiv g^{-1}dg\,, \qquad g \in SU(2)\,,
\] 
the metric of squashed spheres in three dimensions is given by 
\begin{eqnarray}
ds^{2} 
&=&-\frac{L^{2}}{2}\left[
{\rm Tr}\left(J^2\right)
-2C\left({\rm Tr}\left[T^{3}J\right]\right)^{2} \right]\,.  
\label{squashed}
\end{eqnarray}
The deformation parameter $C$ is a real constant supposed to be $C > -1$\,. 
When $C=0$\,, the metric (\ref{squashed}) is reduced to that of round $S^3$ with radius $L$\,. 

\medskip 

For $C \neq 0$\,, the $S^3$ isometry $SO(4)=SU(2)_{\rm L} \times SU(2)_{\rm R}$ is broken to 
$SU(2)_{\rm L} \times U(1)_{\rm R}$\,. 
The $SU(2)_{\rm L}$ transformation is just the left action and
$U(1)_{\rm R}$ transformation is the right action generated by $T^{3}$\,, 
\begin{eqnarray}
g \rightarrow g^L \cdot g \cdot {\rm e}^{-T^{3}\theta}\,. 
\label{trn}
\end{eqnarray}
The infinitesimal forms are 
\begin{eqnarray}
\delta^{L,a}g = \epsilon\,T^{a}g\,, \qquad \delta^{R,3}g = -\epsilon\,gT^{3}\,. 
\label{trans-1}
\end{eqnarray}
The minus sign in the right transformation law comes from the convention in (\ref{trn})\,. 

\medskip 

The classical action of squashed sigma models is given by 
\begin{eqnarray}
S = \int^\infty_{-\infty}\!\!\!dt\int^{\infty}_{-\infty}\!\!\!dx~\eta^{\mu\nu}
\left[{\rm Tr}\left(J_\mu J_\nu\right)-2C{\rm Tr}\left(T^3 J_\mu\right){\rm Tr}\left(T^3 J_\nu\right)\right]\,, \label{action}
\end{eqnarray}
where $x^{\mu}=(t,x)$ with the Lorentzian metric $\eta_{\mu\nu}={\rm diag}(-1,1)$\,. 
We impose the boundary condition that $J_{\mu}$ vanishes at the spatial infinity. 
That is, the group field variable $g(t,x)$ approaches a constant element like\footnote{Seemingly, 
two independent, constant elements are allowed at the two endpoints $x=\pm \infty$\,. 
However, they must be identical by the gauge invariance of the trace of monodromy matrix.} 
\begin{eqnarray}
g(t,x) \rightarrow g_{\infty} \qquad (x\rightarrow\pm\infty)\,. \nonumber 
\end{eqnarray}
The Virasoro constraints are not taken into account, for simplicity. 

\medskip 

The classical equations of motion are 
\begin{eqnarray}
\partial^\mu J_\mu-2C{\rm Tr}(T^3\partial^\mu J_\mu)T^3 -2C{\rm Tr}(T^3J_\mu)\left[J^\mu,T^3\right]=0\,. 
\label{eom} 
\end{eqnarray}
In the squashed sigma models, two infinite-dimensional symmetries 1) quantum affine algebra and 2) Yangian algebra 
are realized and hence two kinds of Lax pairs can be constructed depending on the symmetries. 
That is, there are two descriptions to describe the classical dynamics. 
We shall give a short summary of the two descriptions in the coming two subsections.

\subsection{Trigonometric description}

The one is the trigonometric description related to quantum affine algebra \cite{KMY}. 

\medskip 

With the spectral parameter $\lambda_R$\,, the associated Lax pair is given by 
\cite{FR}\footnote{The study of squashed sigma models has a long history 
and the trigonometric Lax pair was originally constructed by Cherednik \cite{Cherednik}. 
We here use the expression of the Lax pair in \cite{FR}.} 
\begin{eqnarray}
&&L^{R}_{t}(x;\lambda_R)=-\frac{1}{2}\sum_{a=1}^{3}\left[w_a(\alpha + \lambda_R)J^a_+ + w_a(\alpha - \lambda_R)J^a_-\right]T^a\,, \nonumber \\
&&L^{R}_{x}(x;\lambda_R)=-\frac{1}{2}\sum_{a=1}^{3}\left[w_a(\alpha + \lambda_R)J^a_+ - w_a(\alpha - \lambda_R)J^a_-\right]T^a\,, 
\label{right lax} 
\end{eqnarray}
where the following quantities have been introduced,  
\begin{eqnarray}
&& x^{\pm} \equiv \frac{1}{2}(t \pm x)\,, \qquad 
J_{\pm} \equiv J_{t}\pm J_{x}\,, \qquad J_{\mu}^a \equiv -2 {\rm Tr}(T^aJ_{\mu})\,. \nonumber \\ 
&& w_1(\lambda_R) = w_2(\lambda_R) \equiv \frac{\sinh\alpha}{\sinh\lambda_R}\,, 
\quad w_3(\lambda_R) \equiv \frac{\tanh\alpha}{\tanh\lambda_R}\,. 
\nonumber
\end{eqnarray}
The parameter $\alpha$ is related to the squashing parameter $C$ as 
\begin{eqnarray}
i\sqrt{C}=\tanh\alpha\,. \label{alpha}
\end{eqnarray}
Due to the relation (\ref{alpha}) and the reality of $C$\,, $\alpha$ must be pure imaginary for $C>0$ 
and real, up to $i\pi n~(n\in\mathbb{Z})$\,, for $-1<C<0$\,.  
Note that the value of $C$ is automatically restricted to the physical region $C > -1$\,. 
The following zero curvature condition
\begin{eqnarray} 
\bigl[\partial_t-L^{R}_t(x;\lambda_R),\partial_x-L^{R}_x(x;\lambda_R)\bigr]=0 
\label{flatness}
\end{eqnarray}
leads the equations of motion \eqref{eom} and the Maurer-Cartan equation 
$dJ+J\wedge J=0$.

\medskip 

We often discuss the $C \to 0$ limit, 
which corresponds to the $\alpha\to 0$ limit from \eqref{alpha}.  
Before taking the limit, we have to rescale $\lambda_R$ as 
\begin{eqnarray}
\lambda_R = \alpha \tilde{\lambda}_R\,. \label{rescaling}
\end{eqnarray} 
Then the $\alpha \to 0$ limit of (\ref{right lax}) leads to 
the Lax pair of rational type for $SU(2)_{\rm R}$\,.  

\medskip 

\begin{figure}[htbp]
\begin{center}
\begin{tabular}{cc}
\hspace*{-1cm}
\includegraphics[scale=.32]{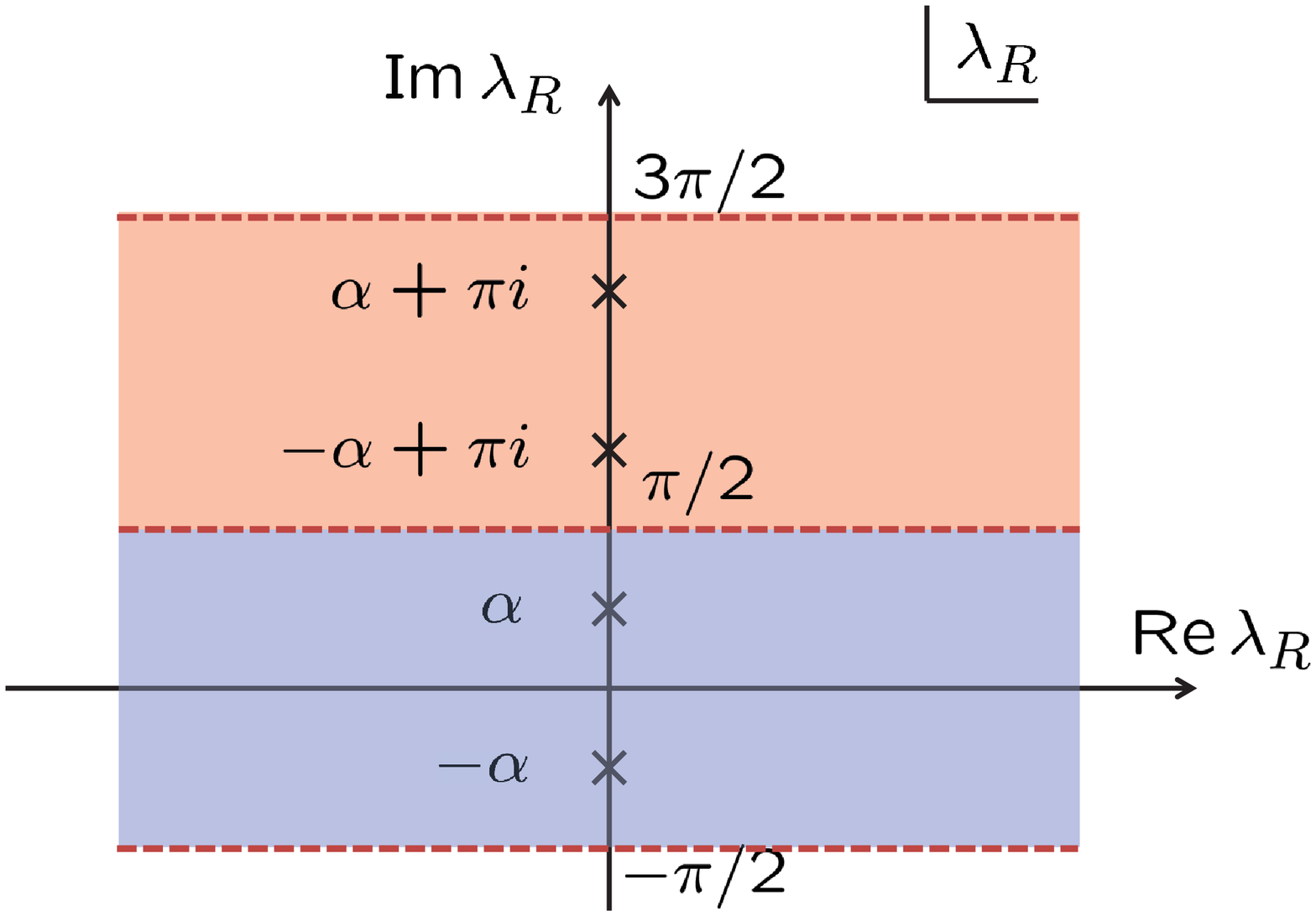}
& 
\hspace*{-1cm}
\includegraphics[scale=.32]{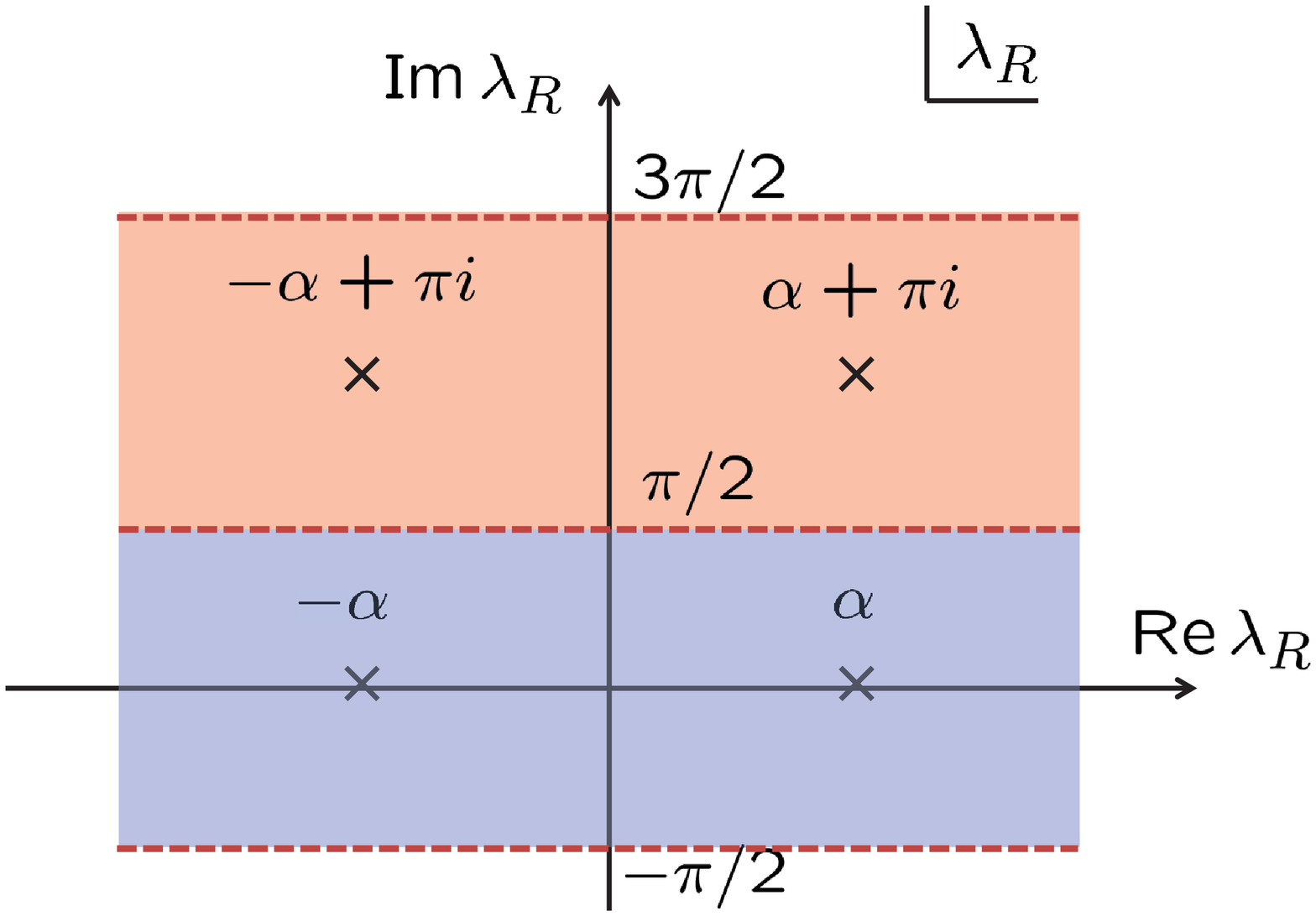} 
\vspace{-1cm}
\\ 
{\footnotesize a) \quad For $C>0$ \qquad }  
& {\footnotesize b) \quad For $-1<C<0$ \qquad }
\end{tabular}
\vspace*{-0.75cm}
\end{center}
\caption{\footnotesize $\lambda_R$ takes values on a cylinder with four punctures. 
\label{strip:fig}}
\end{figure}

It is convenient later to use the light-cone notation like 
\begin{eqnarray}
L^R_\pm(x;\lambda_R) &=& L^R_t(x;\lambda_R) \pm L^R_x(x;\lambda_R) \nonumber \\ 
&=& -\frac{\sinh\alpha}{\sinh\left(\alpha \pm \lambda_R\right)}
\left[T^-J^+_\pm + T^+J^-_\pm + \frac{\cosh\left(\alpha \pm \lambda_R\right)}{\cosh\alpha}T^3J^3_\pm \right]\,,
\label{Lax-tri} 
\end{eqnarray}
where $T^{1,2}$ are recombined into 
\begin{eqnarray}
 T^{\pm} \equiv \frac{1}{\sqrt{2}}\left(T^1 \pm i T^2\right) = T_{\mp}\,.  \nonumber 
\end{eqnarray}
Since the Lax pair given in \eqref{right lax} has the periodicity $2\pi i$ 
with $\lambda_R$ by the definition 
\begin{eqnarray}
L^R_{\pm}(x;\lambda_R) = L^R_{\pm}(x;\lambda_R + 2\pi i)~,
\end{eqnarray}
the spectral parameter $\lambda_R$ can be regarded as living on a cylinder.  
For our convention, the cylinder is parametrized by    
\begin{eqnarray}
 -\frac{\pi}{2} <  {\rm Im}\,\lambda_R  \leq \frac{3}{2}\pi\,. \label{arg}
\end{eqnarray}
The Lax pair (\ref{right lax}) allows $|\lambda_R| = \infty$ but has four poles\footnote{
The number of poles is twice in the relativistic theory in comparison to the non-relativistic case.}, 
\begin{eqnarray}
\lambda_R = \pm \alpha\,, \quad \pm \alpha + \pi i \,. \label{poles}
\end{eqnarray}
Thus the cylinder has four punctures as depicted in Figure \ref{strip:fig}. 

\medskip 

\begin{figure}[b]
\begin{center}
\includegraphics[scale=.5]{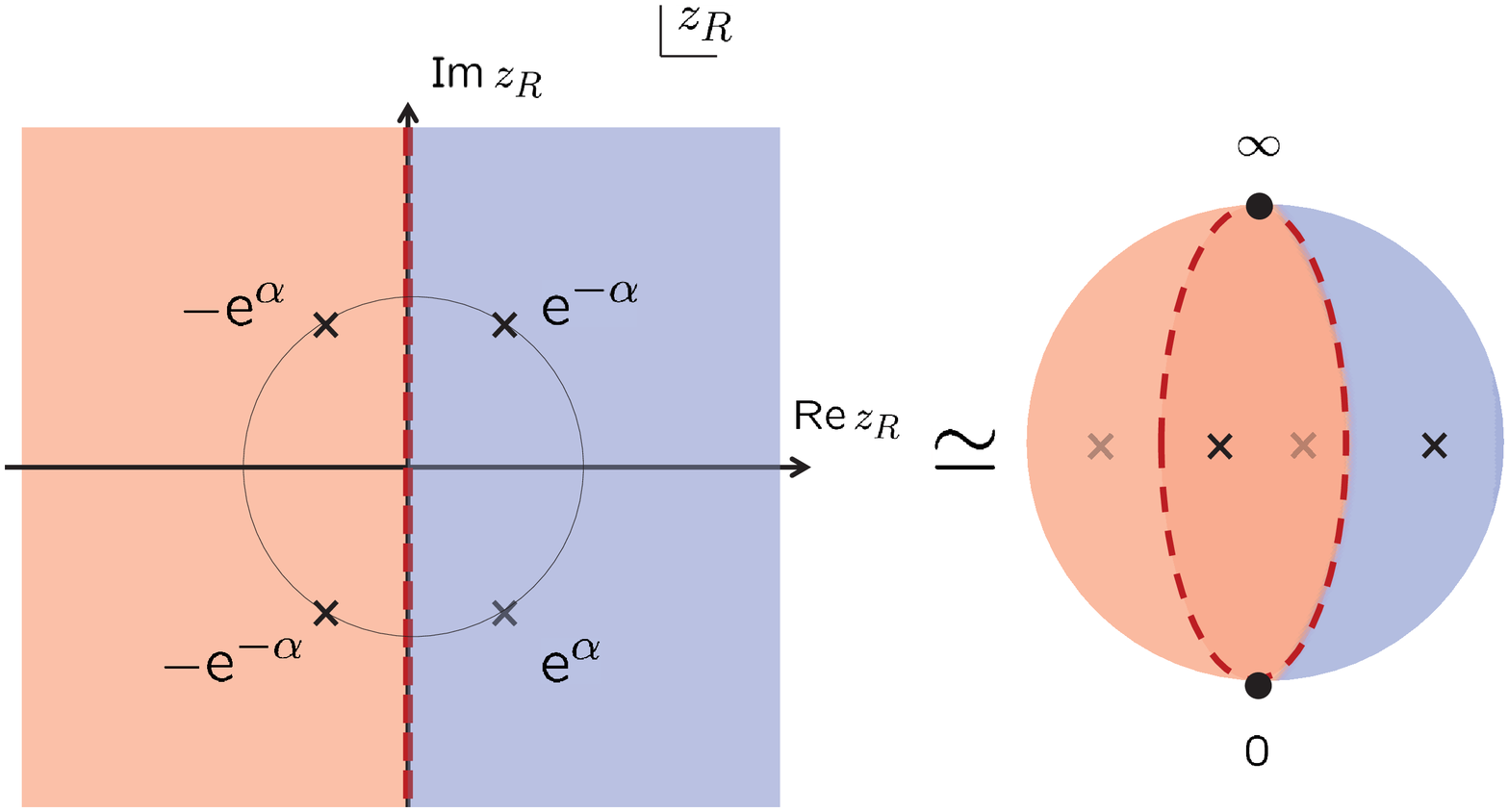}
\vspace*{-2.5cm} 
\end{center}
\caption{\footnotesize The $z_R$-plane for $C>0$ is depicted. It has four punctures but contains $\infty$\,. 
Hence this plane should be regarded as a Riemann sphere with four punctures. \label{z:fig}}
\end{figure}

It is useful to introduce a new parameter defined as 
\begin{eqnarray}
z_R \equiv {\rm e}^{-\lambda_R}\,. \nonumber 
\end{eqnarray}
This maps the $\lambda_R$-cylinder to the $z_R$-plane depicted in Figure \ref{z:fig}. 
According to (\ref{arg})\,, the argument of $z_R$ satisfies 
\begin{eqnarray}
-\frac{3}{2}\pi \leq {\rm arg}(z_R) < \frac{\pi}{2}\,. \nonumber 
\end{eqnarray}

\medskip 

With the spatial component of the Lax pair (\ref{Lax-tri}), the associated monodromy matrix is constructed as 
\begin{eqnarray}
U^R(\lambda_R) = {\rm P}\exp\left[\int^\infty_{-\infty}\!\!\!dx~L^R_x(x;\lambda_R)\right]\,, \label{monodromy-right}
\end{eqnarray}
where the symbol P is the path-ordering. 
Because of the flatness condition \eqref{flatness}, this quantity is conserved 
\[
\frac{d}{dt}U^R(\lambda_R) = 0\,. 
\] 
By expanding $U^R(\lambda_R)$ around $|z_R| < 1$ and $|z_R| > 1$\,, 
the generators of quantum affine algebra are obtained at the classical level \cite{KMY}. 

\medskip 

In the current algebra, non-ultra local terms are contained 
as in the case of principal chiral models \cite{FR}, hence  
there is a subtlety in computing classical $r$-matrix. We follow the $r/s$-matrix formalism \cite{Maillet} 
and show the classical integrability\footnote{The monodromy matrices in \cite{KYhybrid} are computed 
as the retarded monodromy matrices by following \cite{Duncan}. However, it causes a discrepancy 
and we have to follow the $r/s$-matrix formalism \cite{Maillet} as in \cite{KY-summary}. }.  

\medskip 

With the tensor product notation 
\[
\{A\stackrel{\otimes}{,} B\}_{\rm P}\equiv \{A\otimes1,1\otimes B\}_{\rm P}\,,
\] 
the Poisson bracket of the spatial components of the Lax pair is given by 
\begin{eqnarray}
\left\{L^R_x(x;\lambda_R)\stackrel{\otimes}{,}L^R_x(y;\mu_R)\right\}_{\rm P}
&=& \left[r^R(\lambda_R,\mu_R),L^R_x(x;\lambda_R)\otimes 1+1\otimes L^R_x(y;\mu_R)\right]\delta(x-y) \nonumber \\
&& -\left[s^R(\lambda_R,\mu_R),L^R_x(x;\lambda_R)\otimes 1-1\otimes L^R_x(y;\mu_R)\right]\delta(x-y) \nonumber \\
&& -2s^R(\lambda_R,\mu_R)\partial_x\delta(x-y)\,. \nonumber 
\end{eqnarray}
The classical $r$-matrix $r^{L}(\lambda_R,\mu_R)$ and $s$-matrix $s^{L}(\lambda_R,\mu_R)$ are given by \cite{KY-summary} 
\begin{eqnarray}
&&r^{R}(\lambda_R,\mu_R) \equiv \frac{h^R(\lambda_R)+h^R(\mu_R)}{2\sinh{\left(\lambda_R-\mu_R\right)}}\left(T^{+}\otimes T^{-} 
+ T^{-}\otimes T^{+}+\cosh{\left(\lambda_R-\mu_R\right)}T^{3}\otimes T^{3}\right)\,, \nonumber \\
&&s^{R}(\lambda_R,\mu_R) \equiv \frac{h^R(\lambda_R)-h^R(\mu_R)}{2\sinh{\left(\lambda_R-\mu_R\right)}}\left(T^{+}\otimes T^{-} 
+ T^{-}\otimes T^{+} + \cosh{\left(\lambda_R-\mu_R\right)}T^{3}\otimes T^{3}\right)\,, \nonumber 
\end{eqnarray}
where a new function $h^R(\lambda_R)$ is defined as 
\begin{eqnarray}
h^R(\lambda_R) \equiv \frac{\sinh{\alpha}\cosh{\alpha}\sinh^{2}{\lambda_R}}{\sinh{(\alpha-\lambda_R)}\sinh{(\alpha+\lambda_R)}}\,. 
\label{fun}
\end{eqnarray}

\medskip 

It is easy to show the extended classical Yang-Baxter equation is satisfied,
\begin{eqnarray}
&&\left[(r+s)^{R}_{13}(\lambda,\nu),(r-s)^{R}_{12}(\lambda,\mu)\right]
+\left[(r+s)^{R}_{23}(\mu,\nu),(r+s)^{R}_{12}(\lambda,\mu)\right] \nonumber \\
&& \qquad +\left[(r+s)^{R}_{23}(\mu,\nu),(r+s)^{R}_{13}(\lambda,\nu)\right]=0\,, 
\label{YB}
\end{eqnarray}
where the subscripts denote the vector spaces on which the $r$/$s$-matrices act. 

\medskip 

Finally we comment on the pole structure of the $r/s$-matrices. There are two kinds of poles there. 
The first is the four poles of $h^R(\lambda_R)$ given in (\ref{fun}) 
that exactly agrees with those of the Lax pair in (\ref{right lax})\,. 
The second is the two poles coming from the factor $1/\sinh(\lambda_R - \mu_R)$ in the $r$-matrix, 
$\lambda_R = \mu_R$ and $\lambda_R = \mu_R +\pi i$ (or $\lambda_R = \mu_R -\pi i$)\,, depending on the location of $\mu_R$\,. 
Note that there is no pole of the second kind in the $s$-matrix. 
In order for the $r/s$-matrices to satisfy the Yang-Baxter equation (\ref{YB}), the detail form in (\ref{fun}) is irrelevant. 
Therefore we distinguish the class of $r$-matrix in terms of the number of poles in the $r$-matrix apart from the pole coming from 
the scalar function (\ref{fun}). This classification is the same as the one in \cite{BD}. 
According to this criterion, the $r$-matrix in the present case is of trigonometric type.

\subsection{Rational description}

The other is the rational description, which has been developed in \cite{KY}\,, based on the Yangian algebra. 

\medskip 
 
Constructing the Lax pairs in this description, we use the improved currents,
\begin{eqnarray}
 j_{\mu}^{L_\pm} 
= gJ_\mu g^{-1} -2C{\rm Tr}(T^3J_\mu)gT^3g^{-1} \mp\sqrt{C}\epsilon_{\mu\nu}\partial^\nu(gT^3g^{-1})\,. 
\label{improved}
\end{eqnarray}
The anti-symmetric tensor $\epsilon_{\mu\nu}$ is normalized with $\epsilon_{tx}=1$\,. 
The third term is the improvement term so that the currents (\ref{improved}) satisfy 
the flatness condition \cite{KY} 
\begin{eqnarray}
\epsilon^{\mu\nu}\left(\partial_\mu j^{L_\pm}_\nu-j^{L_\pm}_\mu j^{L_\pm}_\nu\right) = 0\,. 
\label{flat-left+}
\end{eqnarray}
There are two types of currents depending on the sign of the improvement term 
and the subscript of $L_{\pm}$ in \eqref{improved} denotes it. 

\medskip 

It is worth noting that, with the improved currents (\ref{improved})\,, 
the classical action (\ref{action}) can be rewritten into a simple, dipole-like form, 
\begin{eqnarray}
 S = \frac{1}{1+C} \int^\infty_{-\infty}\!\!\!dt\int^{\infty}_{-\infty}\!\!\!dx\,
\eta^{\mu\nu}\,{\rm Tr}(j_{\mu}^{L_+}j_{\nu}^{L_{-}})\,. 
\end{eqnarray}
We have not realized the advantage of this expression so far, but it looks very suggestive. 

\medskip 

With the improved currents (\ref{improved}), two kinds of Lax pairs are constructed. 

\medskip 

The one is a Lax pair represented by $j_{\mu}^{L_+}$\,, 
\begin{eqnarray}
L^{L_+}_t(x;\lambda_{L_+}) &\equiv& \frac{1}{1-\lambda_{L_+}^2}\left(j^{L_+}_t - \lambda_{L_+}j^{L_+}_x\right)\,, 
\nonumber \\
L^{L_+}_x(x;\lambda_{L_+}) &\equiv& \frac{1}{1-\lambda_{L_+}^2}\left(j^{L_+}_x - \lambda_{L_+}j^{L_+}_t\right)\,, \nonumber \\
L^{L_+}_\pm(x;\lambda_{L_+}) &\equiv& L^{L_+}_t(x;\lambda_{L_+}) \pm L^{L_+}_x(x;\lambda_{L_+}) 
= \frac{1}{1\pm\lambda_{L_+}}j^{L_+}_\pm  \nonumber \\ 
&=& 
\frac{1}{1\pm \lambda_{L_+}}  g \left( J_\pm  
- 2C\, {\rm Tr}\left(T^3J_\pm\right) T^3  
\mp \sqrt{C}\, \left[J_\pm,T^3\right] 
\right)g^{-1}\,. 
\end{eqnarray}
The terms including $\sqrt{C}$ come from the improvement term. 
The spectral parameter $\lambda_{L_+}$ takes values on a Riemann sphere 
except the poles $\lambda_{L_+} = \pm 1$\,, namely a two punctured Riemann sphere. 
It is noted that the zero curvature condition
\begin{eqnarray} 
\bigl[\partial_t-L^{L_+}_t(x;\lambda_{L_+}),\partial_x-L^{L_+}_x(x;\lambda_{L_+})\bigr]=0 
\label{flatness_left+}
\end{eqnarray}
also leads the equations of motion \eqref{eom} and the flatness condition (\ref{flat-left+})\,.

\medskip 

The associated monodromy matrix 
\begin{eqnarray}
U^{L_+}(\lambda_{L_+}) = {\rm P}\exp\left[\int^\infty_{-\infty}\!\!\!dx~L^{L_+}_x(x;\lambda_{L_+})\right]\, 
\label{monodromy-left+}
\end{eqnarray}
is also conserved due to the zero curvature condition \eqref{flatness_left+} as
\[
\frac{d}{dt}U^{L_+}(\lambda_{L_+}) = 0\,. 
\]
The generators of Yangian are obtained by expanding this monodromy matrix 
around $\lambda_{L_+}=\infty$ \cite{KY}. 
The Poisson bracket of the spatial components of the Lax pair leads to the $r/s$-matrices, 
\begin{eqnarray}
r^L(\lambda_{L_+},\mu_{L_+}) &=& \frac{h^L(\lambda_{L_+})+h^L(\mu_{L_+})}{2\left(\lambda_{L_+}-\mu_{L_+}\right)}
\left(T^+\otimes T^- + T^-\otimes T^+ + T^3\otimes T^3\right)\,, \nonumber \\
s^L(\lambda_{L_+},\mu_{L_+}) &=& \frac{h^L(\lambda_{L_+})-h^L(\mu_{L_+})}{2\left(\lambda_{L_+}-\mu_{L_+}\right)}
\left(T^+\otimes T^- + T^-\otimes T^+ + T^3\otimes T^3\right)\,, \nonumber 
\end{eqnarray} 
where a scalar function $h^L(\lambda_{L_+})$ is defined as 
\begin{eqnarray}
h^L(\lambda_{L_+}) \equiv \frac{C+\lambda_{L_+}^2}{1-\lambda_{L_+}^2}\,.
\end{eqnarray}
The $r$-matrix function has a single pole apart from the poles of $h^L(\lambda_{L_+})$\,. 
Hence the $r$/$s$-matrices are of rational type in the sense of \cite{BD}. 
They satisfy the extended Yang-Baxter equation (\ref{YB})\,. 

\medskip 

The other Lax pair with $j_{\mu}^{L_-}$ is given by  
\begin{eqnarray}
L^{L_-}_t(x;\lambda_{L_-}) &\equiv& \frac{1}{1-\lambda_{L_-}^2}\left(j^{L_-}_t - \lambda_{L_-}j^{L_-}_x\right)\,, \nonumber \\ 
L^{L_-}_x(x;\lambda_{L_-}) &\equiv& \frac{1}{1-\lambda_{L_-}^2}\left(j^{L_-}_x - \lambda_{L_-}j^{L_-}_t\right)\,, \nonumber \\
L^{L_-}_\pm(x;\lambda_{L_-}) &\equiv& L^{L_-}_t(x;\lambda_{L_-}) \pm L^{L_-}_x(x;\lambda_{L_-}) 
= \frac{1}{1\pm\lambda_{L_-}}j^{L_-}_\pm  \nonumber \\ 
&=& 
\frac{1}{1\pm \lambda_{L_-}}  g \left( J_\pm  
- 2C\, {\rm Tr}\left(T^3J_\pm\right) T^3  
\pm \sqrt{C}\, \left[J_\pm,T^3\right] 
\right)g^{-1}\,. 
\end{eqnarray}
The spectral parameter $\lambda_{L_-}$ is independent of $\lambda_{L_+}$ and 
$\lambda_{L_-}$ takes values on another Riemann sphere with two punctures.  
The zero curvature condition is given by 
\begin{eqnarray} 
\bigl[\partial_t-L^{L_-}_t(x;\lambda_{L_-}),\partial_x-L^{L_-}_x(x;\lambda_{L_-})\bigr]=0. 
\label{flatness_left-}
\end{eqnarray}
This Lax pair also leads to the equations of motion (\ref{eom}) and the flatness condition (\ref{flat-left+})\,. 

\medskip 

The monodromy matrix is constructed as 
\begin{eqnarray}
U^{L_-}(\lambda_{L_-}) = {\rm P}\exp\left[\int^\infty_{-\infty}\!\!\!dx~L^{L_-}_x(x;\lambda_{L_-})\right]\,.
\label{monodromy-left-}
\end{eqnarray}
Similarly, this is also a conserved quantity 
because of the zero curvature condition \eqref{flatness_left-} 
and the Poisson bracket of the spatial components of the Lax pair leads to 
the $r/s$-matrices of rational class. 
The resulting $r/s$-matrices are the same as those of $L^{L_+}_x(x;\lambda_{L_+})$\,, 
up to the spectral parameters. It is a convincing result because the $r/s$-matrices depend 
only on $C$\,, not on $\sqrt{C}$\,. 

\medskip 

In the following, we will argue that the monodromy matrices introduced 
in \eqref{monodromy-right}, \eqref{monodromy-left+} and \eqref{monodromy-left-} 
are gauge-equivalent 
under a certain relation of spectral parameters and the rescalings of $sl(2)$ generators.

\section{The trigonometric/rational correspondence}

In order to discuss the direct relations among monodromy matrices, we would like to 
see the correspondence between the trigonometric and rational descriptions 
by expanding the monodromy matrices around some points. 
The data on the expansion points is enough to determine the relation of spectral parameters. 
The necessity of the rescaling of $sl(2)$ generators is anticipated in comparison to the level 
structure of quantum affine algebra.

\subsection{Expansions of $U^R(\lambda_R)$}

Let us consider expanding the monodromy matrix $U^R(\lambda_R)$ in (\ref{monodromy-right}) around some points.

\medskip 

The first is the expansion around $|z_R|<1$ and $|z_R|>1$\,. In these regimes,   
$U^R(\lambda_R)$ is expanded like \cite{KMY}  
\begin{eqnarray}
U^R(\lambda_R) &=& {\rm e}^{\bar{u}_0}\exp\left[\sum_{n=1}^\infty z_R^n\,
\bar{u}_n\right] 
\qquad (|z_R|<1)\,, 
\nonumber \\
U^R(\lambda_R) &=& {\rm e}^{u_0}\exp\left[\sum_{n=1}^\infty z_R^{-n}\,
u_n\right] 
\qquad (|z_R|>1)\,. 
\nonumber 
\end{eqnarray}
Here $u_n$ and $\bar{u}_m$~$(n,m=0,1,\ldots)$ consist of conserved charges. 
For example, the first two of them are  
\begin{eqnarray}
u_0&=&-\bar{u}_0=i\gamma T^3 Q^{R,3}_{(0)}\,, \qquad \gamma \equiv \frac{\sqrt{C}}{1+C}\,, \\
u_1&=&2i\gamma\left(T^- {\rm e}^{\gamma Q^{R,3}_{(0)}/2}Q^{R,+}_{(1)}
+T^+ {\rm e}^{-\gamma Q^{R,3}_{(0)}/2}\widetilde{Q}^{R,-}_{(1)}\right)\,, \nonumber \\
\bar{u}_1&=&-2i\gamma\left(T^+ {\rm e}^{\gamma Q^{R,3}_{(0)}/2}Q^{R,-}_{(1)}
+T^- {\rm e}^{-\gamma Q^{R,3}_{(0)}/2}\widetilde{Q}^{R,+}_{(1)}\right)\,. \nonumber
\end{eqnarray}
The conserved charges  $Q_{(0)}^{R,3}$, $Q_{(1)}^{R,\pm}$ 
and $\widetilde{Q}_{(1)}^{R,\pm}$ precisely generate a quantum affine algebra $U_q(\widehat{sl(2)}_{\rm R})$ 
in the sense of Drinfeld's first realization 
\cite{Drinfeld}. The expressions of the charges are given in \cite{KMY}.\footnote{abc} 
Note that $\gamma$ is related to a $q$-deformation parameter in $U_q(sl(2)_{\rm R})$ 
\cite{Drinfeld,Jimbo} 
through the relation $q \equiv {\rm e}^{\gamma}$ \cite{KYhybrid}.

\medskip 

The second is the expansion around $\lambda_R=0$\,. 
The spatial component of the Lax pair is expanded around $\lambda_R=0$ as  
\begin{eqnarray}
L^R_x(x;\lambda_R) 
&=& -J_x +\lambda_R\left(-\frac{i}{\sqrt{C}}\,J_t + 2i\sqrt{C}\,{\rm Tr}(T^3J_t)T^3\right) \nonumber \\
&& + \lambda_R^2\left[\left(\frac{1}{C}+\frac{1}{2}\right)J_x-\frac{1}{2}{\rm Tr}(T^3J_x)T^3\right] 
+ {\mathcal O}(\lambda_R^3)\,.
\end{eqnarray}
This leads to the expansion of $U^R(\lambda_R)$ around $\lambda_R=0$\,,  
\begin{eqnarray}
U^R(\lambda_R)&=&g_\infty^{-1}\cdot \exp\left[\sum_{n=0}^{\infty}
\left(-\frac{i\lambda_R}{\sqrt{C}}\right)^{n+1} Q^L_{(n)} \right] \cdot g_\infty\,.
\label{ex-Yangian}
\end{eqnarray}
Here $Q_{(n)}^L~(n=0,1,\ldots)$ are the conserved charges because $U^R(\lambda_R)$ is a conserved quantity.  
The first two charges $Q_{(0)}^L$ and $Q_{(1)}^L$ generate the $SU(2)_{\rm L}$ Yangian in the sense of 
Drinfeld's first realization \cite{Drinfeld}. 

\medskip 

Before mentioning $Q_{(0)}^L$ and $Q_{(1)}^L$\,, we have to detail the construction of the Yangian generators.   
By using the improved $SU(2)_{\rm L}$ currents (\ref{improved})\,, 
two kinds of the generators can be constructed as  
\begin{eqnarray}
Q^{L_{\pm}}_{(0)} &=& \int^\infty_{-\infty}\!\!\!dx~j^{L_\pm}_t(x)\,, \nonumber \\
Q^{L_{\pm}}_{(1)} &=& \frac{1}{2}\int^\infty_{-\infty}\!\!\!dx\!\!
\int^\infty_{-\infty}\!\!\!dy~\epsilon(x-y)\,j^{L_\pm}_t(x)j^{L_\pm}_t(y) 
- \int^\infty_{-\infty}\!\!\!dx~j^{L_\pm}_x(x)\,,
\end{eqnarray}
where the signature function $\epsilon(x-y) \equiv \theta(x-y)-\theta(y-x)$ and $\theta(x)$ is a step function. 
Note that the concrete expressions do not depend on $\sqrt{C}$ and hence it is shown that 
\begin{eqnarray}
Q^{L_{+}}_{(0)} = Q^{L_{-}}_{(0)} \qquad 
Q^{L_{+}}_{(1)} = Q^{L_{-}}_{(1)}\,.
\end{eqnarray} 
This is the case for higher conserved charges.  
Thus either of $Q^{L_{+}}_{(n)}$ and $Q^{L_{-}}_{(n)}$ may be taken as $Q^{L}_{(n)}$ in (\ref{ex-Yangian})\,.  
For later discussion, we choose $Q^{L_{+}}_{(n)}$ as $Q^{L}_{(n)}$ here. 

\medskip 

It is quite non-trivial that the $SU(2)_L$ Yangian generators have been reproduced 
by expanding $U^R(\lambda_R)$ around $\lambda_R =0$\,, 
because $U^R(\lambda_R)$ leads to the trigonometric $r/s$-matrices 
while the Yangian is closely related to the rational class. 
Conversely, we show that a quantum affine algebra is reproduced 
by expanding $U^{L_\pm}(\lambda_{L_{\pm}})$ in the next subsection. 

\medskip 

Finally, let us consider the expansion around $\lambda_R=\pi i$\,. 
It also provides the $SU(2)_{\rm L}$ Yangian generators, basically because 
the Lax pair is invariant under the shift of $\lambda_R$ by $\pi i$\,, up to the sign flipping of $T^\pm$\,,  
\begin{eqnarray}
L^R_\pm(x;\lambda_R+\pi i) &=& 
-\frac{\sinh\alpha}{\sinh\left(\alpha \pm \lambda_R\right)}
\left[-T^-J^+_\pm -T^+J^-_\pm +\frac{\cosh\left(\alpha \pm \lambda_R\right)}{\cosh\alpha}T^3J^3_\pm \right]\,. 
\label{flip} 
\end{eqnarray}
This sign flipping is closely related to the rescalings of $sl(2)$ generators discussed later. 
The Yangian charges in this expansion should be identified with $Q^{L_{-}}_{(n)}$\,, according to the choice 
in the expansion around $\lambda_R=0$\,. The reason to assign the charges in this way will be clarified later.

\subsection{Expansions of $U^{L_{\pm}}(\lambda_{L_{\pm}})$} 

We then consider the expansions of $U^{L_+}(\lambda_{L_+})$ and $U^{L_-}(\lambda_{L_-})$ 
around some points. 

\subsubsection*{Expanding $U^{L_+}(\lambda_{L_+})$}

The first is the expansion of $U^{L_+}(\lambda_{L_+})$ around $\lambda_{L_+}=\infty$\,, 
where the $SU(2)_{\rm L}$ Yangian generators are obtained as \cite{KY,KYhybrid} 
\begin{eqnarray}
U^{L_+}(\lambda_{L_+}) = \exp\left[\,\sum_{n=0}^\infty\, \lambda_{L_+}^{-n-1}\,Q^L_{(n)}\,\right]\,.  
\label{ra-Yangian}
\end{eqnarray}
The charges $Q^L_{(n)}$ here are $Q^{L_+}_{(n)}$ by definition, 
and the expansion in (\ref{ra-Yangian}) exactly agrees with the expansion of $U^R(\lambda_R)$ around $\lambda_R=0$\,. 

\medskip 

Next let us consider the expansion around $\lambda_{L_+}=\pm i\sqrt{C}$\,. 
It is convenient to introduce infinitesimal parameters $\epsilon^+_{(\pm)}$ as 
\[
\epsilon^+_{(\pm)} \equiv \lambda_{L_+} \mp i\sqrt{C}\,. 
\]
The expansions of $L^{L_+}_x(x;\lambda_{L_+})$ with respect to $\epsilon^+_{(\pm)}$
are given by, respectively,   
\begin{eqnarray}
L^{L_+}_x(x;\lambda_{L_+}) 
&=& gJ_xg^{-1} +g\left[\mp i\sqrt{C}T^3J^3_t\mp\frac{2i\sqrt{C}}{1+C}T^\pm\left(J^\mp_t\mp i\sqrt{C}J^\mp_x\right)\right]g^{-1} 
\label{left expansion} \\
&& - \frac{\epsilon^+_{(\pm)}}{1+C}g\Biggl[T^\pm\left(\frac{1-C}{1+C}\left(J^\mp_t\mp i\sqrt{C}J^\mp_x\right)\mp\frac{2i\sqrt{C}}{1+C}\left(J^\mp_x\mp i\sqrt{C}J^\mp_t\right)\right) \Biggr. \nonumber \\
&& \Biggl. +T^\mp\left(J^\pm_t\mp i\sqrt{C}J^\pm_x\right)+T^3\left((1-C)J^3_t\mp 2i\sqrt{C}J^3_x\right)\Biggr]g^{-1} +{\mathcal O}\left((\epsilon^+_{(\pm)})^2\right). 
\nonumber
\end{eqnarray}
It would be helpful to introduce the following identity 
\begin{eqnarray}
&&{\rm P}\exp\left[\int^\alpha_\beta dx\left(T^3\partial_x\phi^3+T^+L^-_x+T^-L^+_x\right)(x)\right] \nonumber \\
&=&{\rm e}^{T^3\phi^3(\alpha)}{\rm P}\exp\left[\int^\alpha_\beta dx\left(T^+{\rm e}^{+i\phi^3}L^-_x+T^-{\rm e}^{-i\phi^3}L^+_x\right)(x)\right]{\rm e}^{-T^3\phi^3(\beta)}\,. 
\label{monodromy identity} 
\end{eqnarray} 
This identity is used in the following step. 

\medskip 

The expansion (\ref{left expansion}) and the identity (\ref{monodromy identity}) lead to 
the monodromy matrix $U^{L_+}(\lambda_{L_+})$ in terms of $\epsilon_{(+)}^+$\,,  
\begin{eqnarray}
U^{L_+}(\lambda_{L_+})&=&g_\infty\cdot \bar{v}^+_{(0)}\exp\left[\sum_{n=1}^\infty\left(-\frac{\epsilon^+_{(+)}}{1+C}\right)^n\bar{v}^+_{(n)}\right] \cdot g_\infty^{-1}\,, \label{left monodromy expansion isqrtc}
\end{eqnarray}
where $\bar{v}^+_{(n)}~(n=0,1,\ldots)$ consist of the conserved charges. The first two of them are represented by 
$Q^{R,3}_{(0)}$, $Q^{R,-}_{(1)}$ and $\widetilde{Q}^{R,+}_{(1)}$ as follows: 
\begin{eqnarray}
\bar{v}^+_{(0)}&=&{\rm e}^{-i\gamma T^3 Q^{R,3}_{(0)}}
\exp\left[-2i\gamma T^+{\rm e}^{\gamma Q^{R,3}_{(0)}/2}Q^{R,-}_{(1)}\right]\,, \nonumber \\
\bar{v}^+_{(1)}&=&T^-{\rm e}^{-\gamma Q^{R,3}_{(0)}/2}\widetilde{Q}^{R,+}_{(1)}
-\gamma T^3\left[\bar{Q}^{R,3}_{(2)}-Q^{R,-}_{(1)}\widetilde{Q}^{R,+}_{(1)}\right] \nonumber \\
&&-\gamma^2 T^+{\rm e}^{\gamma Q^{R,3}_{(0)}/2}\left[Q^{R,-}_{(3)}-\bar{Q}^{R,3}_{(2)}Q^{R,-}_{(1)} 
+ \frac{2}{3}\left(Q^{R,-}_{(1)}\right)^2\widetilde{Q}^{R,+}_{(1)}\right]\,. 
\end{eqnarray}
Similarly, the expansion in terms of $\epsilon_{(-)}^-$  is given by 
\begin{eqnarray}
U^{L_+}(\lambda_{L_+})&=&g_\infty\cdot v^+_{(0)}\exp\left[\sum_{n=1}^\infty\left(-\frac{\epsilon^+_{(-)}}{1+C}\right)^nv^+_{(n)}\right] \cdot g_\infty^{-1}\,,  
\end{eqnarray}
where $v^+_{(n)}~(n=0,1,\ldots)$ also consist of the conserved charges and the first two are expressed 
with $Q^{R,3}_{(0)}$, $Q^{R,+}_{(1)}$ and $\widetilde{Q}^{R,-}_{(1)}$ like 
\begin{eqnarray}
v^+_{(0)}&=&{\rm e}^{i\gamma T^3 Q^{R,3}_{(0)}}\exp\left[2i\gamma T^-{\rm e}^{\gamma Q^{R,3}_{(0)}/2}Q^{R,+}_{(1)}\right]\,, \nonumber \\
v^+_{(1)}&=&T^+{\rm e}^{-\gamma Q^{R,3}_{(0)}/2}\widetilde{Q}^{R,-}_{(1)}
-\gamma T^3\left[Q^{R,3}_{(2)}-Q^{R,+}_{(1)}\widetilde{Q}^{R,-}_{(1)}\right] \nonumber \\
&&-\gamma^2 T^-{\rm e}^{\gamma Q^{R,3}_{(0)}/2}\left[Q^{R,+}_{(3)}-Q^{R,3}_{(2)}Q^{R,+}_{(1)}+\frac{2}{3}\left(Q^{R,+}_{(1)}\right)^2\widetilde{Q}^{R,-}_{(1)}\right]\,. 
\end{eqnarray}
In summary, all of the generators of quantum affine algebra have been obtained by expanding $U^{L_+}(\lambda_{L_+})$ 
around $\lambda_{L_{+}}=\pm i \sqrt{C}$\,. This result is also far from trivial 
because $U^{L_+}(\lambda_{L_+})$ yields the rational $r/s$-matrices while 
quantum affine algebra is associated with the trigonometric class.

\subsubsection*{Expanding $U^{L_-}(\lambda_{L_-})$}

It is a turn to discuss the expansions of $U^{L_-}(\lambda_{L_-})$\,. 
We first consider he expansion around $\lambda_{L_-}=\infty$\,, 
where the $SU(2)_{\rm L}$ Yangian generators are obtained as \cite{KY,KYhybrid} 
\begin{eqnarray}
U^{L_-}(\lambda_{L_-})=\exp\left[\sum_{n=0}^\infty (\lambda_{L_-})^{-n-1}Q^L_{(n)}\right]\,. 
\label{ra-Yangian2}
\end{eqnarray}
The charges obtained here are $Q_{(n)}^{L_-}$ by definition, 
and the expansion in (\ref{ra-Yangian2}) exactly agrees with the expansion of $U^R(\lambda_R)$ around $\lambda_R=\pi i$\,. 

\medskip 

Then let us consider the expansion around $\lambda_{L_-}=\pm i\sqrt{C}$\,. 
It is convenient to introduce infinitesimal parameters 
\begin{eqnarray}
\epsilon_{(\pm)}^{-} \equiv \lambda_{L_-} \mp i\sqrt{C}\,. 
\end{eqnarray}
The charges $Q^{R,3}_{(0)}$, $Q^{R,-}_{(1)}$ and $\widetilde{Q}^{R,+}_{(1)}$ are obtained from the expansion in terms of $\epsilon_{(+)}^-$ like 
\begin{eqnarray}
U^{L_-}(\lambda_{L_-})&=&g_\infty\cdot v^-_{(0)}\exp\left[\sum_{n=1}^\infty\left(-\frac{\epsilon^-_{(+)}}{1+C}\right)^nv^-_{(n)}\right] \cdot g_\infty^{-1}\,, \nonumber 
\end{eqnarray}
where $v^-_{(n)}~(n=0,1,\ldots)$ consist of the conserved charges. The first two are given by 
\begin{eqnarray}
v^-_{(0)}&=&{\rm e}^{-i\gamma T^3 Q^{R,3}_{(0)}}\exp\left[-2i\gamma T^-{\rm e}^{-\gamma Q^{R,3}_{(0)}/2}\widetilde{Q}^{R,+}_{(1)}\right]\,, \nonumber \\
v^-_{(1)}&=&T^+{\rm e}^{\gamma Q^{R,3}_{(0)}/2}Q^{R,-}_{(1)}
-\gamma T^3\left[\bar{Q}^{R,3}_{(2)}+\widetilde{Q}^{R,+}_{(1)}Q^{R,-}_{(1)}\right] \nonumber \\
&&-\gamma^2 T^-{\rm e}^{-\gamma Q^{R,3}_{(0)}/2}\left[\widetilde{Q}^{R,+}_{(3)}+\bar{Q}^{R,3}_{(2)}\widetilde{Q}^{R,+}_{(1)}+\frac{2}{3}\left(\widetilde{Q}^{R,+}_{(1)}\right)^2Q^{R,-}_{(1)}\right]\,.
\end{eqnarray}
The remaining charges $Q^{R,3}_{(0)}$, $Q^{R,+}_{(1)}$ and $\widetilde{Q}^{R,-}_{(1)}$ come from the expansion in terms of $\epsilon_{(-)}^-$ as 
\begin{eqnarray}
U^{L_-}(\lambda_{L_-})&=&g_\infty \cdot \bar{v}^-_{(0)}\exp\left[\sum_{n=1}^\infty\left(-\frac{\epsilon^-_{(-)}}{1+C}\right)^n\bar{v}^-_{(n)}\right] \cdot g_\infty^{-1}\,, \nonumber 
\end{eqnarray}
where $\bar{v}^-_{(n)}~(n=0,1,\ldots)$ consist of the conserved charges. The first two are 
\begin{eqnarray}
\bar{v}^-_{(0)}&=&{\rm e}^{i\gamma T^3 Q^{R,3}_{(0)}}\exp\left[2i\gamma T^+{\rm e}^{-\gamma Q^{R,3}_{(0)}/2}\widetilde{Q}^{R,-}_{(1)}\right]\,, \nonumber \\
\bar{v}^-_{(1)}&=&T^-{\rm e}^{\gamma Q^{R,3}_{(0)}/2}Q^{R,+}_{(1)}
-\gamma T^3\left[Q^{R,3}_{(2)}+\widetilde{Q}^{R,-}_{(1)}Q^{R,+}_{(1)}\right] \nonumber \\
&&-\gamma^2 T^+{\rm e}^{-\gamma Q^{R,3}_{(0)}/2}\left[\widetilde{Q}^{R,-}_{(3)}+Q^{R,3}_{(2)}\widetilde{Q}^{R,-}_{(1)}+\frac{2}{3}\left(\widetilde{Q}^{R,-}_{(1)}\right)^2Q^{R,+}_{(1)}\right]\,. 
\end{eqnarray}

\medskip 

The above results can also be obtained by flipping the sign of $\sqrt{C}$ in the results on $U^{L_+}(\lambda_{L_+})$\,. 
Under the sign flipping, $Q^{R,3}_{(0)}$ is invariant 
while $Q^{R,\pm}_{(1)}$ and $\widetilde{Q}^{R,\pm}_{(1)}$ are mapped each other. 

\bigskip 

Finally the results obtained here are summarized in Table \ref{list:tab}. 

\begin{table}[htbp]
\vspace*{0.5cm}
\begin{center}
\begin{tabular}{c||c|c|c}
\hline 
Charges $\setminus$ Monodromies  & $\qquad U^R(\lambda_R) \qquad$ & $\quad U^{L_+}(\lambda_{L_+}) \quad$  
& $\quad U^{L_-}(\lambda_{L_-}) \quad$ \\
\hline\hline 
$Q^{R,3}_{(0)}\,,Q^{R,-}_{(1)}\,,\widetilde{Q}^{R,+}_{(1)}$ & $0$ & $+i\sqrt{C}$ & $+i\sqrt{C}$ \\
\hline
$Q^{R,3}_{(0)}\,,Q^{R,+}_{(1)}\,,\widetilde{Q}^{R,-}_{(1)}$ & $\infty$ & $-i\sqrt{C}$ & $-i\sqrt{C}$ \\ 
\hline
$Q^{L,a}_{(0)}\,,Q^{L,a}_{(1)}$ & $\pm 1$ & $\infty$ & $\infty$ \\
\hline\hline 
local charges & $\pm {\rm e}^{\alpha}$\,,  $\pm {\rm e}^{-\alpha}$ & $\pm 1$ & $\pm 1$\\
\hline 
\end{tabular}
\caption{\footnotesize The conserved charges and the expansion points of monodromy matrices are listed. 
For quantum affine algebra and Yangian, the charges are denoted in the sense of Drinfeld's first realization. 
The expansion points of $U^R(\lambda_R)$ are described in terms of $z_R$\,. 
\label{list:tab}}
\end{center}
\end{table}

\subsection{The relation of spectral parameters}

Now it is a turn to argue the relation of spectral parameters. 
We have already prepared the data 
enough to completely fix it.

\medskip

We assume the relation of spectral parameters is given by a M\"obius transformation. 
Taking account of the correspondence in Table \ref{list:tab}, the M\"obius transformation, 
which relates the expansion points giving the same conserved charges in each description,
is uniquely determined as follows, 
\begin{eqnarray}
z_R^2 = \frac{\lambda_L - i \sqrt{C}}{\lambda_L + i\sqrt{C}} 
\qquad (z_R \equiv {\rm e}^{-\lambda^R})
\,,
\label{map}
\end{eqnarray}
where $\lambda_L~(=\lambda_{L_+}~\mbox{or}~\lambda_{L_-})$\,. 
As we will show in section 4, it is noted that the map \eqref{map} is valid  
not only on some particular expansion points but also on the whole region of the spectral parameters. 
In checking the correspondence of local charges, it is helpful to use the formula,    
\[
\tanh^{-1}x = \frac{1}{2}\log\frac{1+x}{1-x}\,. 
\]

\medskip 

We should be careful for the parameter range of $z_R$\,. 
The relation (\ref{map}) contains the square of $z_R$ and hence two Riemann spheres of $\lambda_L$ 
are basically necessary so that $z_R$ is represented by a single-valued function of $\lambda_{L}$\,. 
Each regime of $\lambda_{L_+}$ and $\lambda_{L_-}$ has already been fixed on a single Riemann sphere 
with two punctures from consistency of the Lax pair in the rational description, 
hence it is not possible to use only either of them.  
Thus it is necessary to use both $\lambda_{L_+}$ and 
$\lambda_{L_-}$\,. After all, $z_R$ is expressed as 
\begin{eqnarray}
z_R = \left\{
\begin{array}{cc}
\displaystyle \quad 
\left(\frac{\lambda_{L_+} - i \sqrt{C}}{\lambda_{L_+} + i\sqrt{C}}\right)^{1/2} & \qquad ({\rm Re}\,z_R >0) \\ 
\displaystyle \quad 
-\left(\frac{\lambda_{L_-} - i \sqrt{C}}{\lambda_{L_-} + i\sqrt{C}}\right)^{1/2} & \qquad ({\rm Re}\,z_R <0)  \\  
\end{array}
\right. \,. \label{map-2}
\end{eqnarray}
This assignment of $\lambda_{L_+}$ and $\lambda_{L_-}$ is compatible with that of $SU(2)_{\rm L}$ Yangian generators.

\begin{figure}[htbp]
\begin{center}
\includegraphics[scale=.6]{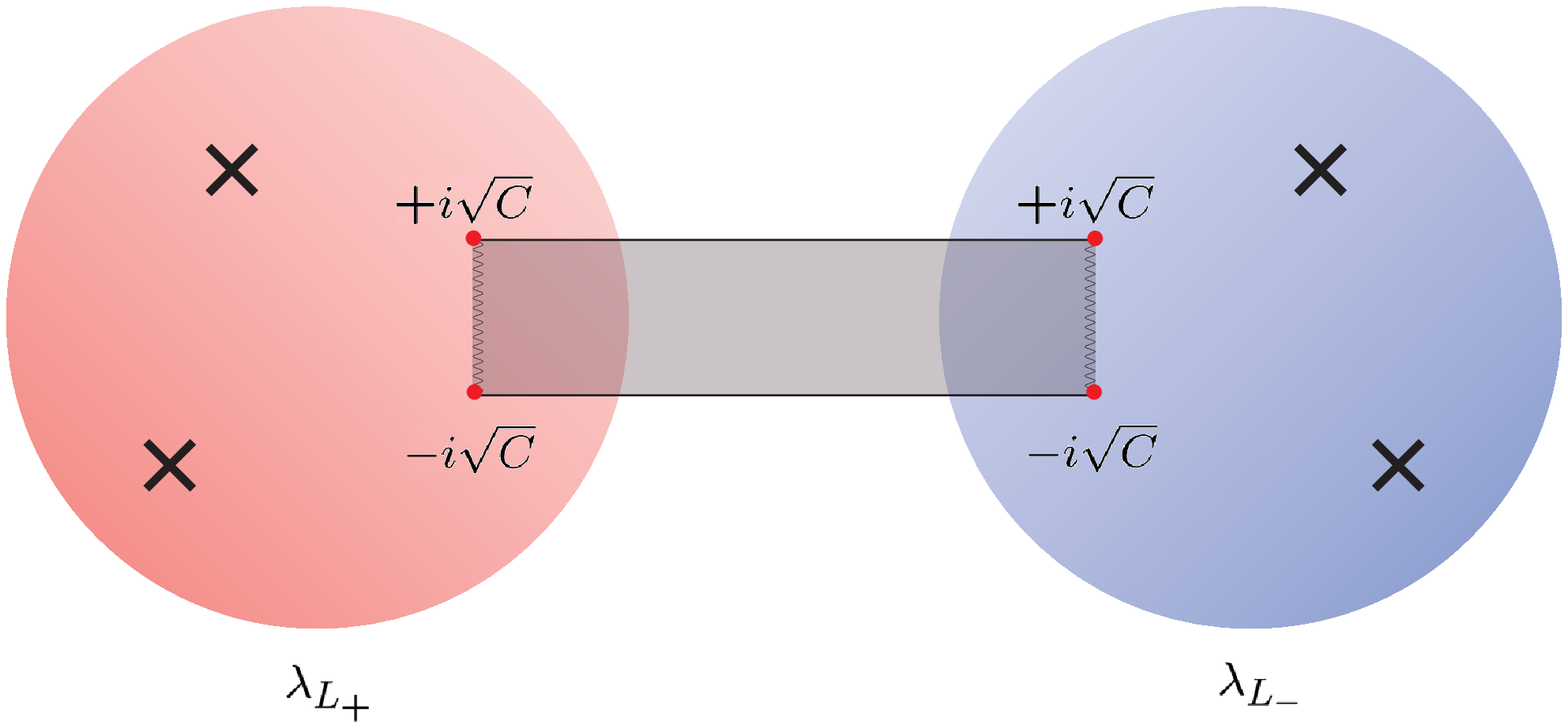}
\vspace*{-5.5cm}
\end{center}
\caption{\footnotesize The $\lambda_{L_{\pm}}$-spheres are joined on the cut between $\pm i \sqrt{C}~(C>0)$\,. 
The constructed Riemann surface is mapped to the Riemann sphere in $z_R$ depicted in Figure \ref{z:fig}. 
\label{lambdaL:fig}}
\end{figure}

\medskip 

In the map (\ref{map-2})\,, there is a cut between $+i\sqrt{C}$ and $-i\sqrt{C}$ on each of the Riemann spheres with $\lambda_{L_+}$ 
and $\lambda_{L_-}$\,, and the two Riemann spheres are joined there as depicted in Fig.\,\ref{lambdaL:fig}. 
Then this cut corresponds to the imaginary axis of $z_R$\,. In order to see this correspondence, 
let us rewrite the relation (\ref{map}) as 
\begin{eqnarray}
\lambda_L=i\sqrt{C}\,\frac{1+z_R^2}{1-z_R^2}\,, \label{map-inv}
\end{eqnarray}
and parametrize the imaginary axis of $z_R$ as 
\begin{eqnarray}
z_R= \pm i\,{\rm e}^{-\xi} \qquad \left(\,\xi\in{\mathbb R}\,\right)\,. \label{axis}
\end{eqnarray}
According to the map (\ref{map-inv})\,, the imaginary axis (\ref{axis}) is represented by 
\begin{eqnarray}
\lambda_L=i\sqrt{C}\tanh\xi  \qquad \left(\,\xi\in{\mathbb R}\,\right) 
\end{eqnarray}
on the $\lambda_{L_{\pm}}$-spheres. This is nothing but the cut in the map (\ref{map-2})\,. 
More precisely, depending on the sign of $C$\,, it is written as 
\begin{eqnarray}
\begin{array}{ccc}
\lambda_L=iy\,, & \quad -\sqrt{C}<y<\sqrt{C} & \qquad (C>0)\,,   \\
\lambda_L=x\,, & \quad -\sqrt{|C|}<x< \sqrt{|C|} & \qquad (-1<C<0)\,.
\end{array} 
\nonumber 
\end{eqnarray}
Thus the resulting Riemann surface described by $\lambda_{L_+}$ and $\lambda_{L_-}$ 
is mapped to the Riemann sphere with $z_R$\,, each other. The number of poles is preserved under the map.

\subsection{The expansions of $U^{L_{\pm}}(\lambda_{L_{\pm}})$~: revisited} 

It is worth reconsidering the expansions of monodromy matrices with the relation (\ref{map}). 

\medskip 

As a concrete example, we will concentrate on the two expansions, 
\begin{center}
\begin{tabular}{rclc}
i) & \quad $U^{L_+}(\lambda_{L_+})$ & around & $\lambda_{L_+} =i\sqrt{C}$\,, \\
ii) & \quad $U^R(\lambda_R)$  & around & $z_R=0$\,.  
\end{tabular}
\end{center}
With the relation (\ref{map})\,, the expansion parameter 
$\epsilon_{(+)}^+$ in the case i) is rewritten as 
\begin{eqnarray}
 \epsilon_{(+)}^+ = \frac{2i\sqrt{C}\,z_R^2}{1-z_R^2}\,. \label{eps}
\end{eqnarray}
Since $|\epsilon_{(+)}^+| \ll 1$\,, $z_R$ is infinitesimal. 
Hence $\epsilon_{(+)}^+$ can be expressed as a power series in $z_R$\,.  
 
\medskip  
 
With the relation (\ref{eps})\,, the expansion in (\ref{left monodromy expansion isqrtc}) can be rewritten as 
\begin{eqnarray}
&&g_\infty^{-1}\cdot U^{L_+}(\lambda_{L_+})\cdot g_\infty \nonumber \\
&=&{\rm e}^{-i\gamma T^3 Q^{R,3}_{(0)}}
\exp\left[-2i\gamma T^+{\rm e}^{\gamma Q^{R,3}_{(0)}/2}Q^{R,-}_{(1)}\right] \nonumber \\
&&\times\left\{1-\frac{2i\sqrt{C}z_R^2}{1+C}\left[T^-{\rm e}^{-\gamma Q^{R,3}_{(0)}/2}\widetilde{Q}^{R,+}_{(1)}
-\gamma T^3\left[\bar{Q}^{R,3}_{(2)}-Q^{R,-}_{(1)}\widetilde{Q}^{R,+}_{(1)}\right] \right.\right. \nonumber \\
&&\left.\left. -\gamma^2 T^+{\rm e}^{\gamma Q^{R,3}_{(0)}/2}\left[Q^{R,-}_{(3)}-\bar{Q}^{R,3}_{(2)}Q^{R,-}_{(1)} 
+ \frac{2}{3}\left(Q^{R,-}_{(1)}\right)^2\widetilde{Q}^{R,+}_{(1)}\right]\right] 
+ {\mathcal O}(z_R^4) \right\}\,. \label{ex} 
\end{eqnarray}
Notice that the rescaling of $sl(2)$ generators 
\begin{eqnarray}
T^{\pm} ~\to~ {\rm e}^{\mp \lambda_R}\,T^{\pm}\,, 
\label{notice}
\end{eqnarray}
makes the expansion in (\ref{ex}) into a significant form, 
\begin{eqnarray}
&&g_\infty^{-1}\cdot U^{L_+}(\lambda_{L_+})\cdot g_\infty \nonumber \\
&=&{\rm e}^{-i\gamma T^3 Q^{R,3}_{(0)}}
\left[1-2i\gamma z_R T^+{\rm e}^{\gamma Q^{R,3}_{(0)}/2}Q^{R,-}_{(1)}\right] \nonumber \\
&&\times\left\{1-2i\gamma z_R^2\left[z_R^{-1}T^-{\rm e}^{-\gamma Q^{R,3}_{(0)}/2}\widetilde{Q}^{R,+}_{(1)}
-\gamma T^3\left[\bar{Q}^{R,3}_{(2)}-Q^{R,-}_{(1)}\widetilde{Q}^{R,+}_{(1)}\right] \right.\right. \nonumber \\
&&\left.\left. -\gamma^2 z_RT^+{\rm e}^{\gamma Q^{R,3}_{(0)}/2}\left[Q^{R,-}_{(3)}-\bar{Q}^{R,3}_{(2)}Q^{R,-}_{(1)} 
+ \frac{2}{3}\left(Q^{R,-}_{(1)}\right)^2\widetilde{Q}^{R,+}_{(1)}\right]\right] \right\} + {\mathcal O}(z_R^2) \nonumber \\
&=&{\rm e}^{-i\gamma T^3 Q^{R,3}_{(0)}}
\left\{1
-2i\gamma z_R\left[ T^+{\rm e}^{\gamma Q^{R,3}_{(0)}/2}Q^{R,-}_{(1)}+T^-{\rm e}^{-\gamma Q^{R,3}_{(0)}/2}\widetilde{Q}^{R,+}_{(1)}\right]
\right\} +{\mathcal O}(z_R^2)\,. \nonumber 
\end{eqnarray}
This is nothing but the expansion in the case ii). That is, if the rescaling (\ref{notice}) is taken into account, 
then the expansion in $\epsilon_{(+)}^+$ can be regarded as the one in $z_R$\,. 
The rescaling (\ref{notice}) is just an isomorphism of the $sl(2)$ algebra, hence it does not mean any modifications of the system.  

\medskip 
 
Similarly, the expansion of $U^{L_+}(\lambda_{L_+})$ around $\lambda_{L_+}=-i\sqrt{C}$ 
agrees with that of $U^R(\lambda_R)$ around $z_R=\infty$ under the relation (\ref{map}) 
with the rescaling (\ref{notice}). 
In addition, the expansion of $U^{L_-}(\lambda_{L_-})$ around $\lambda_{L_{-}}= i\sqrt{C}$ $(-i\sqrt{C})$ 
agrees with that of $U^R(\lambda_R)$ around $z_R=0$ $(\infty)$\,, respectively, if we take another rescaling 
\begin{eqnarray}
T^{\pm} ~ \rightarrow ~  {\rm e}^{\pm \lambda_R}\, T^{\pm}\,. 
\end{eqnarray}

\medskip 

From these agreements, one may anticipate that the rescalings of $sl(2)$ generators
\begin{eqnarray}
&& T^{\pm} ~\to ~{\rm e}^{\mp \lambda_R}\,T^{\pm}  \qquad \mbox{for} \quad U^{L_+}(\lambda_{L_{+}})\,, 
\label{rescaling+} \\ 
&& T^{\pm} ~\to ~{\rm e}^{\pm \lambda_R}\, T^{\pm} \qquad \mbox{for} \quad U^{L_-}(\lambda_{L_{-}})\,. 
\label{rescaling-}
\end{eqnarray}
would be an important key in arguing the equivalence of monodromy matrices. 
Indeed this is the case. The rescalings will play an essential role in the next section.  
Note that the rescalings (\ref{rescaling+}) and (\ref{rescaling-}) are compatible with the sign flipping in (\ref{flip})\,, 
because the shift of $\lambda_R$\,, 
\[
\lambda_R \quad \to \quad \lambda_R + \pi\, i
\] 
flips the sign of $T^{\pm}$ after taking the rescalings (\ref{rescaling+}) and (\ref{rescaling-})\,.

\section{Gauge equivalence of monodromy matrices}

Let us consider the gauge-equivalence of monodromy matrices $U^{R}(\lambda_R)$ 
and $U^{L_{\pm}}(\lambda_{L_{\pm}})$ under the parameter relation (\ref{map}) 
and the rescalings (\ref{rescaling+}) and (\ref{rescaling-}). 

\subsection{Gauge equivalence of monodromy matrices:~$C=0$}

First of all, as a warming-up, we shall consider the $C=0$ case. 
This is nothing but the case of $SU(2)$ principal chiral model 
and its classical integrability is well studied \cite{Luscher1,Luscher2,BIZZ,Bernard-Yangian,MacKay} 
(For a comprehensive book, see \cite{AAR}). 

\medskip 

On the one hand, the Lax pair in terms of the right-invariant current
\[
j^L_\mu \equiv \partial_\mu g\cdot g^{-1}=g\,J_\mu\, g^{-1}\,,
\]
is given by
\begin{eqnarray}
L^L_\pm(x;\lambda_L) = \frac{1}{1\pm\lambda_L}j^L_\pm\,, \label{flat-left}
\end{eqnarray}
where the light-cone components are defined as 
\begin{eqnarray}
L^L_t(x;\lambda_L) \equiv \frac{1}{2}\left[L^L_+(x;\lambda_L) + L^L_-(x;\lambda_L)\right]\,, \quad 
L^L_x(x;\lambda_L) \equiv \frac{1}{2}\left[L^L_+(x;\lambda_L) - L^L_-(x;\lambda_L)\right]\,. \nonumber 
\end{eqnarray}
On the other hand, the Lax pair in terms of the left-invariant current 
\[
j^R_\mu \equiv -g^{-1}\partial_\mu g=-J_\mu\,, 
\]
is given by 
\begin{eqnarray}
L^R_\pm(x;\lambda_R) = \frac{1}{1\pm\lambda_R}j^R_\pm\,, \label{flat-right}
\end{eqnarray}
where the light-components are defined as 
\begin{eqnarray}
L^R_t(x;\lambda_R) \equiv \frac{1}{2}\left[L^R_+(x;\lambda_R) + L^R_-(x;\lambda_R)\right]\,, \quad 
L^R_x(x;\lambda_R) \equiv \frac{1}{2}\left[L^R_+(x;\lambda_R) - L^R_-(x;\lambda_R)\right]\,. \nonumber 
\end{eqnarray}
Then we may introduce monodromy matrices for the Lax pairs (\ref{flat-left}) and (\ref{flat-right}) like 
\begin{eqnarray}
U^L(\lambda_L) &=& {\rm P}\exp\left[\int^\infty_{-\infty}\!\!\!dxL^L_x(x;\lambda_L)\right]\,, \\ 
U^R(\lambda_R) &=& {\rm P}\exp\left[\int^\infty_{-\infty}\!\!\!dxL^R_x(x;\lambda_R)\right]\,.
\end{eqnarray}

\medskip 

From now on, we will show that the two Lax pairs (\ref{flat-left}) and (\ref{flat-right}) are 
gauge-equivalent under the identification of $\lambda_L$ and $\lambda_R$ with  
\begin{eqnarray}
\lambda_L=\frac{1}{\lambda_R}\,. 
\label{relation C=0}
\end{eqnarray}
First of all, let us perform the gauge transformation for the $L^L_{\pm}(x;\lambda_L)$\,. 
Since the Lax pair is transformed as a gauge field, the transformation law is give by 
\begin{eqnarray}
\Bigl[L^L_\pm(x;\lambda_L)\Bigr]^g \equiv g^{-1}L^L_\pm(x;\lambda_L)g - g^{-1}\partial_\pm g  
= -\frac{\pm\lambda_L}{1\pm\lambda_L}J_\pm\,. 
\end{eqnarray}
Using the relation (\ref{relation C=0})\,, we can show that 
\begin{eqnarray}
\Bigl[L^L_\pm(x;\lambda_L)\Bigr]^g
= L^R_\pm(x;\lambda_R)\,. 
\end{eqnarray}
With this relation, we obtain the following formula for covariant derivatives,
\begin{eqnarray}
g^{-1}\left[\partial_\mu-L^L_\mu(x;\lambda_L)\right]g = \partial_\mu-L^R_\mu(x;\lambda_R)\,. 
\end{eqnarray}
Thus the transformation law of monodromy matrices is given by 
\begin{eqnarray}
g_\infty^{-1}\cdot U^L(\lambda_L)\cdot g_\infty=U^R(\lambda_R)\,,
\end{eqnarray}
and we have shown that $U^L(\lambda_L)$ is gauge-equivalent to $U^R(\lambda_R)$ 
under the identification (\ref{relation C=0})\,. 

\medskip 

It would be interesting to see the gauge-equivalence at $r/s$-matrix level. 
The $r/s$-matrices are derived from the following Poisson brackets,   
\begin{eqnarray}
\left\{L^L_x(x;\lambda_L)\stackrel{\otimes}{,}L^L_x(y;\mu_L)\right\}_{\rm P}
&=& \left[r^L(\lambda_L,\mu_L),L^L_x(x;\lambda_L)\otimes 1+1\otimes L^L_x(y;\mu_L)\right]\delta(x-y) \nonumber \\
&& -\left[s^L(\lambda_L,\mu_L),L^L_x(x;\lambda_L)\otimes 1-1\otimes L^L_x(y;\mu_L)\right]\delta(x-y) \nonumber \\
&& -2s^L(\lambda_L,\mu_L)\partial_x\delta(x-y)\,, \nonumber \\
\left\{L^R_x(x;\lambda_R)\stackrel{\otimes}{,}L^R_x(y;\mu_R)\right\}_{\rm P}
&=& \left[r^R(\lambda_R,\mu_R),L^R_x(x;\lambda_R)\otimes 1+1\otimes L^R_x(y;\mu_R)\right]\delta(x-y) \nonumber \\
&& -\left[s^R(\lambda_R,\mu_R),L^R_x(x;\lambda_R)\otimes 1-1\otimes L^R_x(y;\mu_R)\right]\delta(x-y) \nonumber \\
&& -2s^R(\lambda_R,\mu_R)\partial_x\delta(x-y)\,. \nonumber 
\end{eqnarray}
The classical $r/s$-matrices are the following: 
\begin{eqnarray}
&&r^L(\lambda_L,\mu_L) = \frac{h(\lambda_L)+h(\mu_L)}{2\left(\lambda_L-\mu_L\right)}
\left(T^+\otimes T^- + T^-\otimes T^+ + T^3\otimes T^3\right)\,, \nonumber \\
&&s^L(\lambda_L,\mu_L) = \frac{h(\lambda_L)-h(\mu_L)}{2\left(\lambda_L-\mu_L\right)}
\left(T^+\otimes T^- + T^-\otimes T^+ + T^3\otimes T^3\right)\,, \nonumber \\
&&r^R(\lambda_R,\mu_R) = \frac{h(\lambda_R)+h(\mu_R)}{2\left(\lambda_R-\mu_R\right)}
\left(T^+\otimes T^- + T^-\otimes T^+ + T^3\otimes T^3\right)\,, \nonumber \\
&&s^R(\lambda_R,\mu_R) = \frac{h(\lambda_R)-h(\mu_R)}{2\left(\lambda_R-\mu_R\right)}
\left(T^+\otimes T^- + T^-\otimes T^+ + T^3\otimes T^3\right)\,. 
\end{eqnarray}
Here $h(\lambda)$ is defined as 
\begin{eqnarray}
h(\lambda) \equiv \frac{\lambda^2}{1-\lambda^2}\,.
\end{eqnarray}
It is straightforward to compute the gauge transformation laws of $r/s$-matrices. 
The $r$-matrix is transformed as  
\begin{eqnarray}
\Bigl[r^L(\lambda_L,\mu_L)\delta(x-y)\Bigr]^g &=& g^{-1}(x)\otimes g^{-1}(y)\left[r^L(\lambda_L,\mu_L)\delta(x-y) \right. \nonumber \\ 
&&- \left. \frac{1}{2}\left\{g(x)\stackrel{\otimes}{,}L^L_x(y;\mu_L)\right\}_{\rm P}g^{-1}(x)\otimes 1 
\right. \nonumber \\ && \left. 
-\frac{1}{2}\left\{L^L_x(x;\lambda_L)\stackrel{\otimes}{,}g(y)\right\}_{\rm P}1\otimes g^{-1}(y)\right]g(x)\otimes g(y)\,, \nonumber 
\end{eqnarray}
and the $s$-matrix is transformed as 
\begin{eqnarray}
\Bigl[s^L(\lambda_L,\mu_L)\delta(x-y)\Bigr]^g &=& g^{-1}(x)\otimes g^{-1}(y)\left[s^L(\lambda_L,\mu_L)\delta(x-y) \right. \nonumber \\ 
&& \left. + \frac{1}{2}\left\{g(x)\stackrel{\otimes}{,}L^L_x(y;\mu_L)\right\}_{\rm P}g^{-1}(x)\otimes 1\right. \nonumber \\
&& \left.-\frac{1}{2}\left\{L^L_x(x;\lambda_L)\stackrel{\otimes}{,}g(y)\right\}_{\rm P}1\otimes g^{-1}(y)\right]g(x)\otimes g(y)\,. \nonumber 
\end{eqnarray}
With the Poisson brackets,
\begin{eqnarray}
&&\left\{g(x)\stackrel{\otimes}{,}L^L_x(y;\mu_L)\right\}_{\rm P} 
=\frac{\mu_L}{\mu_L^2-1}\left(T^+\otimes T^- + T^-\otimes T^+ + T^3\otimes T^3 \right)g(x)\otimes 1\delta(x-y)\,, \nonumber \\
&&\left\{L^L_x(x;\lambda_L)\stackrel{\otimes}{,}g(y)\right\}_{\rm P} =
-\frac{\lambda_L}{\lambda_L^2-1}\left(T^+\otimes T^- + T^-\otimes T^+ + T^3\otimes T^3 \right)1\otimes g(y)\delta(x-y)\,. \nonumber 
\end{eqnarray}
the gauge-equivalence of $r/s$-matrices are shown as 
\begin{eqnarray}
\Bigl[r^L(\lambda_L,\mu_L)\Bigr]^g = r^R(\lambda_R,\mu_R)\,, \qquad 
\Bigl[s^L(\lambda_L,\mu_L)\Bigr]^g = s^R(\lambda_R,\mu_R)\,. \nonumber 
\end{eqnarray}
This equivalence still holds even after squashing the target space geometry, 
as we will see in the next subsection. 

\medskip 

Finally we should emphasize the advantage of $r/s$-matrix formalism. If one uses the retarded monodromy 
matrix in \cite{Duncan}, instead of the $r/s$-matrix formalism, then the gauge-equivalence does not hold.

\subsection{Gauge equivalence of monodromy matrices:~$C \neq 0$}

It is a turn to consider the squashed sigma model case with $C \neq 0$\,. 
Here we will show that $U^R(\lambda_R)$ is gauge-equivalent to $U^{L_{\pm}}(\lambda_{L_{\pm}})$ 
under the relation (\ref{map}) and the rescalings (\ref{rescaling+}) and (\ref{rescaling-})\,. 

\medskip 

Let us start from rewriting the Lax pair $L_{\pm}^{L_+}(x;\lambda_{L_+})$ as 
\begin{eqnarray}
&&L^{L_+}_\pm(x;\lambda_{L_+}) \nonumber \\ 
&=& 
\frac{1}{1\pm \lambda_{L_+}}  g \left[ T^+\!\left(1\mp i\sqrt{C}\right)\!J^-_\pm 
\!+\! T^-\!\left(1\pm i\sqrt{C}\right)\!J^+_\pm 
\!+\! T^3\!\left(1+C\right)\!J^3_\pm
\right]g^{-1}\,. \nonumber 
\end{eqnarray} 
As in the case with $C=0$\,, a gauge transformation of it is evaluated as 
\begin{eqnarray}
&& \Bigl[L^{L_+}_\pm(x;\lambda_{L_+})\Bigr]^g 
\equiv g^{-1}L^{L_+}_\pm(x;\lambda_{L_+})g -g^{-1}\partial_\pm g   \nonumber \\
&=& -J_\pm 
+ \frac{1}{1\pm \lambda_{L_+}} \left[ T^+\!\left(1\mp i\sqrt{C}\right)\!J^-_\pm 
\!+\! T^-\!\left(1\pm i\sqrt{C}\right)\!J^+_\pm 
\!+\! T^3\!\left(1+C\right)\!J^3_\pm
\right] \nonumber \\
&=& -\frac{\pm\lambda_{L_+}}{1\pm\lambda_{L_+}}\left[ T^+\!\left(1+\frac{i\sqrt{C}}{\lambda_{L_+}}\right)\!J^-_\pm 
\!+\! T^-\!\left(1-\frac{i\sqrt{C}}{\lambda_{L_+}}\right)\!J^+_\pm 
\!+\! T^3\!\left(1\mp\frac{C}{\lambda_{L_+}}\right)\!J^3_\pm
\right]\,. \nonumber 
\end{eqnarray}
By using the inverse relation of (\ref{map})\,, 
\begin{eqnarray}
\lambda_{L_\pm} = \frac{\tanh\alpha}{\tanh\lambda_{R}}\,, 
\label{relation Cneq0}
\end{eqnarray}
the gauge transformation is further rewritten as 
\begin{eqnarray}
\Bigl[L^{L_+}_\pm(x;\lambda_{L_+})\Bigr]^g 
= -\frac{\sinh\alpha}{\sinh(\alpha\pm\lambda_{R})}
\left[ T^+{\rm e}^{\lambda_{R}}J^-_\pm 
\!+\! T^-{\rm e}^{-\lambda_{R}}J^+_\pm 
\!+\! T^3\frac{\cosh(\alpha\pm\lambda_{R})}{\cosh\alpha}J^3_\pm
\right]\,. \nonumber 
\end{eqnarray}
Thus, up to the rescaling (\ref{rescaling+}), we have shown that 
\begin{eqnarray}
\Bigl[L^{L_+}_\pm(x;\lambda_{L_+})\Bigr]^g \simeq L^{R}_\pm(x;\lambda_{R})\,.
\end{eqnarray}
This relation means the gauge-equivalence of monodromy matrices,
\begin{eqnarray}
g_\infty^{-1}\cdot U^{L_+}(\lambda_{L_+})\cdot g_\infty \simeq U^{R}(\lambda_{R})\,. \label{equiv+}
\end{eqnarray}
Note that only half of the range of $\lambda_R$ is covered by $\lambda_{L_+}$\,,  
as we know from (\ref{map-2})\,.  

\medskip 

The same argument is possible for $U^{L_-}(\lambda_{L_-})$\,. Only the difference is that the rescaling (\ref{rescaling-}) 
has to be used instead of (\ref{rescaling+})\,. Then we obtain that 
\begin{eqnarray}
g_\infty^{-1}\cdot U^{L_-}(\lambda_{L_-})\cdot g_\infty \simeq U^{R}(\lambda_{R})\,. \label{equiv-}
\end{eqnarray}
The remaining range of $\lambda_R$ is covered by $\lambda_-$\,. 
Thus, putting (\ref{equiv+}) and (\ref{equiv-}) together, we have shown that $U^R(\lambda_R)$ is gauge-equivalent to 
$U^{L_{\pm}}(\lambda_{L_\pm})$\,. 

\medskip 

Let us comment on the rescalings (\ref{rescaling+}) and (\ref{rescaling-})\,. They can be expressed 
as a transformation generated by ${\rm e}^{\mp iT^3 \lambda_R}$\,. Then the Lax pair is transformed as   
\begin{eqnarray}
L^{R_\pm}_\mu(x;\lambda_R) = {\rm e}^{\pm iT^3 \lambda_R}\,L^R_\mu(x;\lambda_R)\,{\rm e}^{\mp iT^3 \lambda_R}\,. 
\label{gauge3}
\end{eqnarray}
With this transformation law, the gauge-equivalence of monodromy matrices is represented by a simple form,  
\begin{eqnarray}
\widetilde{g}_\pm^{-1}\cdot U^{L_{\pm}}(\lambda_{L_\pm})\cdot \widetilde{g}_\pm = U^R(\lambda_R)\,, 
\qquad \widetilde{g}_\pm \equiv g_\infty\cdot{\rm e}^{\pm iT^3 \lambda_R}\,.
\end{eqnarray}
Note that $\lambda_R$ is a complex variable and hence the transformation (\ref{gauge3}) is not 
an $SU(2)_{\rm L}$ transformation.  

\medskip 

The next task is to check the gauge-equivalence at the $r/s$-matrix level. 
Recall that the left and right $r/s$-matrices are given by \cite{KY-summary} 
\begin{eqnarray}
&&r^{L_\pm}(\lambda_{L_\pm},\mu_{L_\pm}) = \frac{h^{L}(\lambda_{L_\pm})+h^{L}(\mu_{L_\pm})}{2\left(\lambda_{L_\pm}-\mu_{L_\pm}\right)}
\left(T^+\otimes T^- + T^-\otimes T^+ + T^3\otimes T^3\right)\,, \nonumber \\
&&s^{L_\pm}(\lambda_{L_\pm},\mu_{L_\pm}) = \frac{h^{L}(\lambda_{L_\pm})-h^{L}(\mu_{L_\pm})}{2\left(\lambda_{L_\pm}-\mu_{L_\pm}\right)}
\left(T^+\otimes T^- + T^-\otimes T^+ + T^3\otimes T^3\right)\,, \nonumber \\
&& r^{R}(\lambda_{R},\mu_{R}) = \frac{h^{R}(\lambda_{R})+h^{R}(\mu_{R})}{2\sinh\left(\lambda_{R}-\mu_{R}\right)}
\left(T^+\otimes T^- + T^- \otimes T^+ 
+ \cosh\left(\lambda_{R}-\mu_{R}\right)T^3\otimes T^3\right)\,, \nonumber \\
&&s^{R}(\lambda_{R},\mu_{R}) = \frac{h^{R}(\lambda_{R})-h^{R}(\mu_{R})}{2\sinh\left(\lambda_{R}-\mu_{R}\right)}
\left(T^+\otimes T^- + T^-\otimes T^+ 
+ \cosh\left(\lambda_{R}-\mu_{R}\right)T^3\otimes T^3\right)\,. \nonumber 
\end{eqnarray}
Here scalar functions $h^L(\lambda_L)$ and $h^R(\lambda_R)$ are defined as, respectively,  
\begin{eqnarray}
h^L(\lambda_L) &\equiv& \frac{C+\lambda_L^2}{1-\lambda_L^2}\,, \label{funcs0} \\ 
h^R(\lambda_R) &\equiv& \frac{\sinh\alpha\cosh\alpha\sinh^2\lambda_R}{\sinh(\alpha-\lambda_R)\sinh(\alpha+\lambda_R)}\,. 
\label{funcs}
\end{eqnarray}

Under the gauge transformation, the rational $r/s$-matrices are transformed as 
\begin{eqnarray}
&&\Bigl[r^{L_\pm}(\lambda_{L_\pm},\mu_{L_\pm})\delta(x-y)\Bigr]^g \nonumber \\
&=&g^{-1}(x)\otimes g^{-1}(y)\left[r^{L_\pm}(\lambda_{L_\pm},\mu_{L_\pm})\delta(x-y)
-\frac{1}{2}\left\{g(x)\stackrel{\otimes}{,}L^{L_\pm}_x(y;\mu_{L_\pm})\right\}_{\rm P}g^{-1}(x)\otimes 1\right. \nonumber \\
&&\qquad\qquad\qquad\qquad\left.
-\frac{1}{2}\left\{L^{L_\pm}_x(x;\lambda_{L_\pm})\stackrel{\otimes}{,}g(y)\right\}_{\rm P}1\otimes 
g^{-1}(y)\right]g(x)\otimes g(y) \nonumber \\
&&\Bigl[s^{L_\pm}(\lambda_{L_\pm},\mu_{L_\pm})\delta(x-y)\Bigr]^g \nonumber \\
&=&g^{-1}(x)\otimes g^{-1}(y)\left[s^{L_\pm}(\lambda_{L_\pm},\mu_{L_\pm})\delta(x-y)
+\frac{1}{2}\left\{g(x)\stackrel{\otimes}{,}L^{L_\pm}_x(y;\mu_{L_\pm})\right\}_{\rm P}g^{-1}(x)\otimes 1\right. \nonumber \\
&&\qquad\qquad\qquad\qquad\left.
-\frac{1}{2}\left\{L^{L_\pm}_x(x;\lambda_{L_\pm})\stackrel{\otimes}{,}g(y)\right\}_{\rm P}1\otimes 
g^{-1}(y)\right]g(x)\otimes g(y)\,. \nonumber 
\end{eqnarray}
By using the following Poisson brackets, 
\begin{eqnarray}
&&\left\{g(x)\stackrel{\otimes}{,}L^{L_\pm}_x(y;\mu_L)\right\}_{\rm P} \label{Poisson1}  \\
&=&\left\{\frac{-\mu_L}{1-\mu_L^2}\left(T^+\!\otimes\!T^- + T^-\!\otimes\!T^+ 
+ T^3\!\otimes\!T^3 \right) \right. \nonumber \\
&&\pm\left. \frac{\sqrt{C}}{1-\mu_L^2}\left[T^+\!\otimes\!T^- 
+ T^-\!\otimes\!T^+ + T^3\!\otimes\!T^3,gT^3 g^{-1}(x)\!\otimes\!1\right]
\right\}g(x)\!\otimes\!1\delta(x-y)\,, \nonumber 
\\
&&\left\{L^{L_\pm}_x(x;\lambda_L)\stackrel{\otimes}{,}g(y)\right\}_{\rm P} \label{Poisson2} \\
&=&\left\{\frac{\lambda_L}{1-\lambda_L^2}\left(T^+\!\otimes\!T^- 
+ T^-\!\otimes\!T^+ + T^3\!\otimes\!T^3 \right) \right. \nonumber \\
&&\mp\left. \frac{\sqrt{C}}{1-\lambda_L^2}\left[T^+\!\otimes\!T^- 
+ T^-\!\otimes\!T^+ + T^3\!\otimes\!T^3,1\!\otimes\!gT^3 
g^{-1}(y)\right]\right\}1\!\otimes\!g(y)\delta(x-y)\,, \nonumber
\end{eqnarray}
and the rescalings (\ref{rescaling+}) and (\ref{rescaling-})\,, 
the gauge-equivalence of $r$/$s$-matrices are shown as 
\begin{eqnarray}
&& \Bigl[r^{L_\pm}(\lambda_{L_\pm},\mu_{L_\pm})\Bigr]^g \simeq r^R(\lambda_R,\mu_R)\,, \qquad 
\Bigl[s^{L_\pm}(\lambda_{L_\pm},\mu_{L_\pm})\Bigr]^g \simeq s^R(\lambda_R,\mu_R)\,. 
\nonumber 
\end{eqnarray}

\medskip 

At first glance, it might seem contradictory because the number of poles of $h^L(\lambda_L)$ 
is two and that of $h^R(\lambda_R)$ 
is four. However, the map (\ref{map-2}) means that the range of $\lambda_R$ is divided into the two regions, 
hence the number of poles is also compatible.  
This is the case for the pole of $r$-matrix apart from those in $h^L(\lambda_L)$ and $h^R(\lambda_R)$\,.  
Its number is just one and exactly agrees with the number in either of the rational descriptions.

\medskip 

Finally we should comment on the $C \to 0$ limit. 
The relation (\ref{relation Cneq0}) is reduced to the relation (\ref{relation C=0}) 
in the $C\rightarrow 0$ limit with the rescaling (\ref{rescaling}). 

\subsection{Reduced trigonometric description and integrability}

In the previous argument, one may have noticed the possibility that a couple of the two Lax pairs 
\begin{eqnarray}
&& L^{R_+}_\pm(x;\lambda_{R_+})  \nonumber \\ 
&=& -\frac{\sinh\alpha}{\sinh\left(\alpha \pm \lambda_{R_+}\right)} 
\left[{\rm e}^{-\lambda_{R_+}}T^-J^+_\pm + {\rm e}^{\lambda_{R_+}}T^+J^-_\pm 
+ \frac{\cosh\left(\alpha \pm \lambda_{R_+}\right)}{\cosh\alpha}T^3J^3_\pm \right]\,, \nonumber \\
&&L^{R_-}_\pm(x;\lambda_{R_-}) \nonumber \\ 
&=& -\frac{\sinh\alpha}{\sinh\left(\alpha \pm \lambda_{R_-}\right)} 
\left[{\rm e}^{\lambda_{R_-}}T^-J^+_\pm + {\rm e}^{-\lambda_{R_-}}T^+J^-_\pm 
+ \frac{\cosh\left(\alpha \pm \lambda_{R_-}\right)}{\cosh\alpha}T^3J^3_\pm \right]\,, \label{irr-Lax}
\end{eqnarray}
are available in the trigonometric description, instead of the Lax pair $L^R_{\mu}(x;\lambda_R)$ in (\ref{right lax})\,. 
Now that two spectral parameters $\lambda_{R_{\pm}}$ are contained in the Lax pairs (\ref{irr-Lax}), 
the periodicity of ${\rm Im}\,\lambda_{R_\pm}$ is $\pi$, not $2\pi$~: 
\begin{eqnarray}
L^{R_\pm}_\mu(x;\lambda_{R_\pm}\!+\pi i) = L^{R_\pm}_\mu(x;\lambda_{R_\pm})\,. \label{half}
\end{eqnarray}
This observation implies that the Lax pair (\ref{right lax}) is ``reducible'' in some sense. 
In fact, it is straightforward to check that each of the Lax pairs (\ref{irr-Lax}) leads to 
the identical classical equations of motion (\ref{eom}). 
Hence it really works well as the Lax pair, at least, at the classical level, 
though it is unclear whether it works even at the quantum level or not. 

\medskip 

The two spectral parameters decompose the relation (\ref{map}) into the two relations, 
\begin{eqnarray}
\lambda_{L_{\pm}}=\frac{\tanh\alpha}{\tanh\lambda_{R_{\pm}}}\,. \label{map-re}
\end{eqnarray}
With the relation (\ref{map-re})\,, a gauge-transformation of $L^{L_+}_\pm(x;\lambda_{L_+})$ 
can be shown as 
\begin{eqnarray}
&& \Bigl[L^{L_+}_\pm(x;\lambda_{L_+}) \Bigr]^g
= g^{-1}L^{L_+}_\pm(x;\lambda_{L_+})g - g^{-1}\partial_\pm g  \nonumber \\
&=& -\frac{\pm\lambda_{L_+}}{1\pm\lambda_{L_+}}\left[ T^+\!\left(1+\frac{i\sqrt{C}}{\lambda_{L_+}}\right)\!J^-_\pm 
\!+\! T^-\!\left(1-\frac{i\sqrt{C}}{\lambda_{L_+}}\right)\!J^+_\pm 
\!+\! T^3\!\left(1\mp\frac{C}{\lambda_{L_+}}\right)\!J^3_\pm
\right] \nonumber \\
&=& -\frac{\sinh\alpha}{\sinh(\alpha\pm\lambda_{R_+})}
\left[ T^+{\rm e}^{\lambda_{R_+}}J^-_\pm 
\!+\! T^-{\rm e}^{-\lambda_{R_+}}J^+_\pm 
\!+\! T^3\frac{\cosh(\alpha\pm\lambda_{R_+})}{\cosh\alpha}J^3_\pm
\right] \nonumber \\
&=& L^{R_+}_\pm(x;\lambda_{R_+})\,.
\end{eqnarray}
Thus we have shown the gauge-equivalence as 
\begin{eqnarray}
\Bigl[L^{L_+}_\pm(x;\lambda_{L_+}) \Bigr]^g  = L^{R_+}_\pm(x;\lambda_{R_+})\,, 
\end{eqnarray}
without the rescalings of $sl(2)$ generators. 
The gauge-equivalence of 
$L^{L_-}_\pm(x;\lambda_{L_-})$ and $L^{R_-}_\pm(x;\lambda_{R_-})$ 
can also be shown in the same way. 

\medskip 

To summarize, the monodromy matrices satisfy the relations,   
\begin{eqnarray}
g_\infty^{-1}\cdot U^{L_{\pm}}(\lambda_{L_{\pm}})\cdot g_\infty=U^{R_{\pm}}(\lambda_{R_{\pm}})\,.
\end{eqnarray}

\medskip 

The next is to consider the $r/s$-matrices related to a pair of the Lax pairs (\ref{irr-Lax})\,. 
From the Poisson brackets of the spatial components of the Lax pairs (\ref{irr-Lax})\,, 
similarly, one can read off the $r/s$-matrices,  
\begin{eqnarray}
&&r^{R_{\pm}}(\lambda_{R_{\pm}},\mu_{R_{\pm}}) = \frac{h^{R}(\lambda_{R_{\pm}})+h^{R}(\mu_{R_{\pm}})}{
2\sinh\left(\lambda_{R_{\pm}}-\mu_{R_{\pm}}\right)}
\left({\rm e}^{\pm\left(\lambda_{R_{\pm}}-\mu_{R_{\pm}}\right)}T^+\otimes T^- \right. \nonumber \\ 
&& \hspace{3cm} \left. + {\rm e}^{\mp\left(\lambda_{R_{\pm}}-\mu_{R_{\pm}}\right)}T^-\otimes T^+ 
+ \cosh\left(\lambda_{R_{\pm}}-\mu_{R_{\pm}}\right)T^3\otimes T^3\right)\,, 
\label{r-red} \\
&&s^{R_{\pm}}(\lambda_{R_{\pm}},\mu_{R_{\pm}}) = \frac{h^{R}(\lambda_{R_{\pm}})-h^{R}(\mu_{R_{\pm}})}{
2\sinh\left(\lambda_{R_{\pm}}-\mu_{R_{\pm}}\right)}
\left({\rm e}^{\pm\left(\lambda_{R_{\pm}}-\mu_{R_{\pm}}\right)}T^+\otimes T^- \right. \nonumber \\
&& \hspace{3cm} \left. + {\rm e}^{\mp\left(\lambda_{R_{\pm}}-\mu_{R_{\pm}}\right)}T^-\otimes T^+ 
+ \cosh\left(\lambda_{R_{\pm}}-\mu_{R_{\pm}}\right)T^3\otimes T^3\right)\,. 
\label{s-red}
\end{eqnarray}
Here a scalar function 
$h^R(\lambda_{R_{\pm}})$ is already introduced in (\ref{funcs}).
The $r/s$-matrices satisfy the extended Yang-Baxter equation (\ref{YB}), and the classical integrability has 
also been shown based on the Lax pair (\ref{irr-Lax})\,.

\medskip 

Note that the range of $\lambda_{R_{\pm}}$ is restricted to 
half of the original trigonometric one as in (\ref{half})\,. So the number of poles in $h^R(\lambda_{R_{\pm}})$ 
is just two and it agrees with that in either of the rational descriptions. 
This is the case for the poles of the $r$-matrix apart from the poles in $h^R(\lambda_{R_{\pm}})$ 
and it is just one. Thus the $r$-matrix is really of rational type in the sense of \cite{BD}, 
though it does not look so. 

\medskip 

This can be confirmed by showing that  
the $r/s$-matrices in the rational and reduced trigonometric descriptions 
are related each other by a gauge transformation. 
The Poisson brackets (\ref{Poisson1}) and (\ref{Poisson2}) lead to the transformation laws 
\begin{eqnarray}
\Bigl[r^{L_{\pm}}(\lambda_{L_{\pm}},\mu_{L_{\pm}})\delta(x-y)\Bigr]^g 
= r^{R_{\pm}}(\lambda_{R_{\pm}},\mu_{R_{\pm}})\delta(x-y)\,, \label{tr1} \\
\Bigl[s^{L_{\pm}}(\lambda_{L_{\pm}},\mu_{L_{\pm}})\delta(x-y)\Bigr]^g 
= s^{R_{\pm}}(\lambda_{R_{\pm}},\mu_{R_{\pm}})\delta(x-y)\,, \label{tr2}
\end{eqnarray}
without rescaling the $sl(2)$ generators. The relations (\ref{tr1}) and (\ref{tr2}) confirm that 
the $r/s$-matrices (\ref{r-red}) and (\ref{s-red}) should be regarded as those of rational type.

\medskip 

Thus the trigonometric Lax pair (\ref{right lax}) is really reducible to a pair of the rational Lax pairs (\ref{irr-Lax}) 
applicable to the classical analysis of the squashed sigma models. 
It would be interesting to consider how to interpret the reducibility of the Lax pair (\ref{right lax}), at the quantum level, 
especially in the language of Bethe ansatz \cite{FR,PW,quantum1,quantum2,quantum3}.

\section{Conclusion and discussion}

We have shown the gauge-equivalence of monodromy matrices in the trigonometric and 
rational descriptions under the relation of spectral parameters and the rescalings of $sl(2)$ 
generators. As a result, the trigonometric description has been shown to be equivalent to a pair of 
the rational descriptions. 
That is, the ``global'' equivalence is accurately realized 
even after squashing the target space geometry. 
Moreover, we have found the trigonometric description is reducible to a pair of the 
``reduced'' trigonometric descriptions, each of which is of rational class and works well as the Lax pair at the classical level. 
With this description, the equivalence of monodromy matrices has become very apparent. 

\medskip 

The equivalence implies that a squashed sphere is represented by a pair of round spheres 
as a dipole from the viewpoint of classical integrability. This is equivalent to say that 
a warped AdS$_3$ space is a pair of undeformed AdS$_3$ spaces via a double Wick rotation. 
This dipole-like structure of target space would correspond to that of dipole CFT$_2$ 
in warped AdS$_3$/CFT$_2$ \cite{Guica,SS}. 
It is a challenging issue to try to establish the correspondence in the scenario. 

\medskip 

In this direction the rational description would play an important role based on the ``global'' 
equivalence because a Virasoro symmetry is realized as a reprametrization of the initial data in the solution 
generating techniques, as dicussed in    
\cite{Schwarz,DS,Pope}. This Virasoro symmetry is different from the one coming from 
the classical conformal symmetry of the system. 
Thus we speculate that the former Virasoro algebra and the initial data 
can be related to the quantities in the conjectured dual ``dipole CFT'' \cite{Guica,SS}. 
This scenario might give a successful way to identify the dual CFT 
at the sigma model level, while the asymptotic symmetry analysis at the gravity level 
has not completely succeeded so far. Similarly, the related Kac-Moody algebra can also be discussed 
\cite{Schwarz,DS,Pope}\footnote{It would also be nice to consider the relation to the scenario discussed for KdV equations \cite{BLZ}.}.   
It would also be interesting to seek a direct connection to the theorem recently presented by Hofman and Strominger \cite{HS}. 

\medskip 

It is of importance to look for the purely mathematical formulation of the correspondence between 
a quantum affine algebra and a pair of Yangians, without the sigma model framework. 
Another issue is to consider the RTT relation in light of the correspondence. 
It would be useful to follow the quantum treatment of quantum affine algebra \cite{Bernard} 
and the Bethe ansatz \cite{PW,quantum1,quantum2,quantum3}. Notably, the trigonometric and rational 
S-matrices are contained in the Bethe ansatz. If the equivalence shown here survives quantization, 
the Bethe ansatz may be rewritten into the one consisting of only the rational S-matrices 
but with the same spectrum.

\medskip 

It would also be nice to consider the similar correspondence of monodromy matrices 
in the case of null-warped AdS$_3$\,, where the broken $SL(2)_{\rm R}$ symmetry 
is realized as a $q$-deformed Poincare symmetry \cite{KY-Sch}. Its affine extension has not been clarified yet, 
but, conversely, it may be done by using the gauge-equivalence of monodromy matrices.

\subsection*{Acknowledgments}

We would like to thank S.~Moriyama for useful discussions. 
The work of IK was supported by the Japan Society for the Promotion of Science (JSPS). 
The work of KY was supported by the scientific grants from the Ministry of Education, Culture, Sports, Science 
and Technology (MEXT) of Japan (No.\,22740160). This work was also supported in part by the Grant-in-Aid 
for the Global COE Program ``The Next Generation of Physics, Spun 
from Universality and Emergence'' from MEXT, Japan. 
One of the author TM also would like to thank A.~Molev for valuable discussions and comments.

\end{document}